\PassOptionsToPackage{table}{xcolor}
\documentclass[acmsmall,screen,nonacm]{acmart}

\usepackage{graphicx}
\usepackage{subcaption}
\usepackage{placeins} 
\usepackage{float}     
\usepackage{tcolorbox}
\tcbuselibrary{skins,breakable,listings}
\usepackage{listings}
\usepackage{xcolor}

\lstdefinestyle{code}{
  basicstyle=\ttfamily\small,
  numbers=left, numberstyle=\scriptsize, numbersep=6pt,
  breaklines=true, breakatwhitespace=true,
  frame=single, framerule=0.4pt,
  tabsize=2, showstringspaces=false
}
\lstdefinestyle{cpp}{style=code, language=C++}
\lstdefinestyle{json}{style=code, language=}
\lstdefinestyle{txt}{style=code, language=}

\newtcblisting{codecard}[2][]{%
  enhanced, breakable,
  colback=white, colframe=black!12,
  boxrule=0.6pt, arc=2mm, left=1mm, right=1mm, top=1mm, bottom=1mm,
  title={#2}, fonttitle=\bfseries\footnotesize,
  listing only, listing options={#1}
}

\usepackage{xcolor}

\usepackage{wrapfig}
\usepackage{multirow}
\usepackage{enumitem}
\usepackage{listings}
\usepackage{xspace}
\usepackage{subcaption}

\usepackage{tikz}
\tikzset{>=latex}
\newcommand*\circled[1]{\tikz[baseline=(char.base)]{
            \node[shape=circle,draw,fill=black,text=white,inner
            sep=0.4pt] (char) {\small #1};}}

\usepackage[framemethod=tikz]{mdframed}
\mdfdefinestyle{mpdframe}{
   frametitlebackgroundcolor   =black!15,
    frametitlerule              =true,
    roundcorner                 =3pt,
    middlelinewidth             =1pt,
    skipabove                   =\topskip,
    skipbelow                   =\topskip,
    innermargin                 =0.1cm, 
    outermargin                 =0.1cm,
    innerleftmargin             =0.1cm,
    innerrightmargin            =0cm,
    innertopmargin              =0.1cm,
    innerbottommargin           =0.1cm,
    align=center   
}

\newcommand{\specialcellr}[1]{\begin{tabular}[c]{@{}r@{}}#1\end{tabular}}
\newcommand{\DefMacro}[2]{%
   \expandafter\newcommand\csname rmk-#1\endcsname{#2}%
}
\newcommand{\UseMacro}[1]{\csname rmk-#1\endcsname}

\newcommand{\eg}{e.g.}
\newcommand{\ie}{i.e.}

\newcommand{\Comment}[1]{}

\newcommand{\Space}[1]{}

\newcommand{\MyPara}[1]{\vspace{1pt}\noindent\textbf{#1}.}

\newcommand{\MyItPara}[1]{\noindent\textit{#1}.}

\newcommand{\Code}[1]{{\small\ifmmode{\texttt{#1}}\else$\texttt{#1}$\fi}}
\newcommand{\CodeIn}[1]{{\small\ifmmode{\mathtt{#1}}\else$\mathtt{#1}$\fi}}
\newcommand{\ColorBack}[1]{%
  \begingroup \setlength{\fboxsep}{0pt}
}

\definecolor{gray}{RGB}{211,211,211}

\newcommand{\PaperTitle}{Evaluating the Effectiveness of Coverage-Guided Fuzzing for Testing Deep Learning Library APIs}

\newcommand{\Contrib}[1]{$\star$#1}

\newcommand{\EQ}[1]{RQ#1\xspace}

\newcommand{\sota}{SoTA\xspace}
\newcommand{\cgf}{CGF\xspace}
\newcommand{\tname}{\textsc{FlashFuzz}\xspace}
\newcommand{\acetest}{\textsc{ACETest}\xspace}
\newcommand{\ttfuzz}{\textsc{TitanFuzz}\xspace}
\newcommand{\ffuzz}{\textsc{FreeFuzz}\xspace}
\newcommand{\pathfinder}{\textsc{PathFinder}\xspace}
\newcommand{\dl}{DL\xspace}
\newcommand{\torch}{PyTorch\xspace}
\newcommand{\tf}{TensorFlow\xspace}

\newcommand{\torchvereval}{2.2.0\xspace}
\newcommand{\tfvereval}{2.16.1\xspace}
\newcommand{\torchverwild}{2.7.0\xspace}
\newcommand{\tfverwild}{2.19.0\xspace}

\newcommand{\rqOneTorchTotalAPIs}{1,151\xspace}
\newcommand{\rqOneTorchIntersectACE}{284\xspace}
\newcommand{\rqOneTorchIntersectPathfinder}{529\xspace}
\newcommand{\rqOneTorchIntersectTTFuzz}{469\xspace}

\newcommand{\rqOneTFTotalAPIs}{662\xspace}
\newcommand{\rqOneTFIntersectACE}{519\xspace}
\newcommand{\rqOneTFIntersectPathfinder}{478\xspace}
\newcommand{\rqOneTFIntersectTTFuzz}{643\xspace}

\newcommand{\libfuzzer}{libFuzzer\xspace}
\newcommand{\afl}{AFL++\xspace}

\newcommand{\RepoURL}{\url{https://github.com/ncsu-swat/FlashFuzz}}

\newcommand{\numBugs}{42\xspace}
\newcommand{\numBugsTorch}{20\xspace}
\newcommand{\numBugsTF}{22\xspace}
\newcommand{\numBugsFix}{8\xspace}

\newcommand{\todo}[2][normal]{%
  \ifdefined\showtodos
    \ifstrequal{#1}{high}{\sethlcolor{red}}{}%
    \ifstrequal{#1}{normal}{\sethlcolor{yellow}}{}%
    \ifstrequal{#1}{low}{\sethlcolor{green}}{}%
    {\hl{\textbf{TODO:} #2}}%
    \marginpar{\textcolor{red}{\textbf{TODO}}}%
  \fi
}

\newif\ifshowtodos
\showtodostrue  

\DefMacro{eq-single-performance}{\EQ{1}}
\DefMacro{eq-single-safety}{\EQ{2}}
\DefMacro{eq-evolution-stability}{\EQ{3}}
\DefMacro{eq-evolution-performance}{\EQ{4}}
\DefMacro{eq-evolution-safety}{\EQ{5}}
\DefMacro{Category}{Category}
\DefMacro{PolicyType}{Policy Type}
\DefMacro{Description}{Description}

\newenvironment{packed_itemize}{
\vspace{-0.1ex}
\begin{list}{\labelitemi}{\leftmargin=1.7em}
 \setlength{\itemsep}{0pt} \setlength{\parskip}{0pt} \setlength{\parsep}{0pt} \setlength{\headsep}{0pt} \setlength{\topskip}{0pt} \setlength{\topmargin}{0pt} \setlength{\topsep}{0pt} \setlength{\partopsep}{0pt}
 }{\end{list}
\vspace{-0.1ex}
}

\newenvironment{packed_enumerate}{
\vspace{-0.7ex}
\begin{enumerate}[label=\arabic*.,leftmargin=1.7em]
 \setlength{\itemsep}{0.7pt}
 \setlength{\parskip}{0pt}
 \setlength{\parsep}{0pt}
 \setlength{\headsep}{0pt}
 \setlength{\topskip}{0pt}
 \setlength{\topmargin}{0pt}
 \setlength{\topsep}{0pt}
 \setlength{\partopsep}{0pt}
}{\end{enumerate}
\vspace{-0.7ex}
}

\definecolor{SubtleColor}{rgb}{0,0,.50}

\newcommand{\deltax}[1]{(~#1$\times$)}

\newcommand{\torchDeltaTotalAce}{17.1}
\newcommand{\torchDeltaTotalPathfinder}{4.6}
\newcommand{\torchDeltaTotalTT}{1182.0}

\newcommand{\torchDeltaValidAce}{21.3}
\newcommand{\torchDeltaValidPathfinder}{4.7}
\newcommand{\torchDeltaValidTT}{1271.3}

\newcommand{\torchDeltaRatioAce}{1.3}
\newcommand{\torchDeltaRatioPathfinder}{1.0}
\newcommand{\torchDeltaRatioTT}{1.1}

\newcommand{\tfDeltaTotalAce}{1.0}
\newcommand{\tfDeltaTotalPathfinder}{1.7}
\newcommand{\tfDeltaTotalTT}{535.3}

\newcommand{\tfDeltaValidAce}{2.5}
\newcommand{\tfDeltaValidPathfinder}{9.1}
\newcommand{\tfDeltaValidTT}{909.2}

\newcommand{\tfDeltaRatioAce}{2.5}
\newcommand{\tfDeltaRatioPathfinder}{5.4}
\newcommand{\tfDeltaRatioTT}{1.7}

\newcommand{\covtfffvspf}{212.88\%}
\newcommand{\covtfffvstt}{101.13\%}

\setcopyright{none}
\acmYear{2025}
\acmDOI{}
\acmISBN{}

\begin{document}

\fancyfoot{}

\title{\PaperTitle}

\author{Feiran Qin}
\affiliation{
  \institution{North Carolina State University}
  \city{Raleigh}
  \state{NC}
  \country{USA}
}
\email{fqin2@ncsu.edu}

\author{M. M. Abid Naziri}
\affiliation{
  \institution{North Carolina State University}
  \city{Raleigh}
  \state{NC}
  \country{USA}
}
\email{mnaziri@ncsu.edu}

\author{Hengyu Ai}
\affiliation{
  \institution{ShanghaiTech University}
  \city{Shanghai}
  \country{China}
}
\email{aihy2023@shanghaitech.edu.cn}

\author{Saikat Dutta}
\affiliation{
  \institution{Cornell University}
  \city{Ithaca}
  \state{NY}
  \country{USA}
}
\email{saikatd@cornell.edu}

\author{Marcelo d'Amorim}
\affiliation{
  \institution{North Carolina State University}
  \city{Raleigh}
  \state{NC}
  \country{USA}
}
\email{mdamori@ncsu.edu}

\renewcommand{\shortauthors}{Qin et al.}


\begin{abstract} 
Deep Learning (\dl) libraries (e.g., \torch) provide the core components to
build major AI-enabled applications.\Comment{that are prolific today.}
Finding bugs in these libraries is an important and challenging problem. Prior
approaches have tackled this challenge by performing either API-level fuzzing or
model-level fuzzing. However, these approaches do not use coverage-guidance,
which limits their effectiveness and efficiency. This limitation raises an
intriguing question: can coverage-guided fuzzing (\cgf) -- in particular,
established frameworks like LibFuzzer -- be effectively applied to \dl
libraries, and does it offer meaningful improvements in code coverage, bug
detection, and scalability compared to prior methods?




In this paper, we perform the first in-depth study to answer this question. We
observe that a key challenge in applying \cgf\ to test \dl\ libraries is the
need to create a complete test harness, for each \dl API, that can correctly
transform byte-level inputs generated by the fuzzer into valid inputs that
satisfy the complex input constraints of the API. To address this challenge and
support this study, we propose \tname, a technique that leverages Large Language
Models (LLMs) to automatically synthesize API-level test harnesses by combining
harness templates, domain-specific helper functions, and API documentation. 
%
%
\tname\ uses a feedback-driven strategy to iteratively synthesize and repair
harnesses. Using this approach, \tname can successfully synthesize test
harnesses for \rqOneTorchTotalAPIs and \rqOneTFTotalAPIs APIs of \torch\ and
\tf, respectively. Compared to state-of-the-art fuzzing techniques for \dl\
libraries (namely \acetest, \pathfinder, and \ttfuzz), \tname\ achieves up to
\covtfffvstt-\covtfffvspf{} higher coverage and
\torchDeltaRatioPathfinder{}x-\tfDeltaRatioPathfinder{}x higher validity rate.
Compared to these baselines, \tname\ achieves speedups of 1x-1182x in generating
inputs. Finally, \tname\ has also discovered \numBugs previously unknown bugs in
the latest versions of \torch\ and \tf, of which \numBugsFix have already been
fixed by developers. Thus, our study confirms that \cgf\ can be successfully
applied to test \dl\ libraries, and provides a strong baseline for future testing
approaches.

%
\end{abstract}  


\maketitle
\pagestyle{plain}

\section{Introduction}
\label{sec:intro}
Deep Learning (\dl) libraries, such as \tf~\cite{abadi2016tensorflow} and
\torch~\cite{paszke2019pytorch} are the backbone of modern machine learning
applications. Like any piece of complex software, these libraries
contain bugs~\cite{Jia_ETAL_JSS21,Xie2022DocTer,Wei2022,Chen_ETAL_TOSEM23}.
\emph{Model-based
fuzzing}~\cite{Phan_ETAL_ICSE2019,AUDEE_ASE20,LEMON_FSE20,MUFFIN_ICSE22,Liu_2023,Liu_ETAL_ICSE23}
and \emph{API-level
fuzzing}~\cite{Wei2022,Deng_ETAL_FSE22,Christou_ETAL_USENIX23,Deng_ETAL_ISSTA23,PathFinder_ICSE25}
are two complementary categories of bug finding techniques for \dl\ library testing. 
Model-based fuzzing generates programs representing an
ML model using the APIs from the \dl\ library whereas API-level
fuzzing focuses on generating inputs to individual APIs of the
library. We focus on API-level fuzzing, which is crucial for uncovering bugs that may not manifest at the model level.


Coverage-guided fuzzers (\cgf), such as \afl~\cite{aflnew} and \libfuzzer~\cite{libfuzzer}, have uncovered bugs across diverse domains, including operating systems~\cite{274553}, compilers~\cite{10.1109/ICSE48619.2023.00017}, and network protocols~\cite{boofuzz}. These fuzzers are known for their scalability, coverage-driven input generation, and low setup overhead. However, their application to deep learning libraries remains largely unexplored. This paper reports on a study to assess the effectiveness of \cgf\ for testing \dl\ libraries. This question is important given the limited scalability or coverage of existing techniques. For example, a class of techniques use expensive external tools, such as LLMs (e.g., \ttfuzz~\cite{Deng_ETAL_ISSTA23}) and SMT solvers (e.g., \acetest~\cite{Shi_2023} and \pathfinder~\cite{PathFinder_ICSE25}) to generate inputs and a class of techniques are oblivious to
 complex input constraints in the APIs~(e.g., \ffuzz~\cite{Wei2022}), limiting their
 ability to cover deeper logic in code. 
 To sum up, we hypothesize that \cgf{}s can efficiently explore the input spaces of \dl APIs to uncover bugs. This paper evaluates that hypothesis.
 


An important challenge in
using \cgf\ for API-level testing relates to scaling the creation of
test harnesses to handle APIs with various signatures and input
constraints. To address this challenge and support this study, we propose \tname, a technique that
leverages Large Language Models (LLMs) to automatically synthesize test harnesses from
templates, helper functions, and API documentation. 
Like recent API-level fuzzers (\eg{}, \acetest~\cite{Shi_2023} and
\pathfinder~\cite{PathFinder_ICSE25}), \tname\ builds on the observation that
the majority of the important code of APIs are in the kernel functions,
implemented in C++. Other LLM-based harness generation techniques exist (\eg{},
PromptFuzz~\cite{lyu2024prompt} and CKGFuzzer~\cite{xu2024ckgfuzzer}) but
they are not specialized to \dl\ APIs. Section~\ref{sec:related} elaborates on
related work.


%

\vspace{0.5ex}
\MyPara{Method}~\tname\ wraps \libfuzzer~\cite{libfuzzer} with an automated test
harness synthesis approach. We evaluate \tname\ (\ie{}, the \cgf\ method it
represents) against three \sota\ techniques using the standard metrics from the
literature: coverage, validity ratio, and number of crashes revealed. We select
\acetest~\cite{Shi_2023} and \pathfinder~\cite{PathFinder_ICSE25} as they
analyze kernel implementations, as \tname\ does, and we select
\ttfuzz~\cite{Deng_ETAL_ISSTA23} as it represents a technique that uses Large
Language Models (LLMs) to generate inputs. Furthermore, we conducted 8-hour fuzzing campaigns with the harnesses \tname\ generates to evaluate their ability to find new bugs.

\subsection{Results}~ Our results show that \tname outperforms the comparison
baselines on all metrics. In terms of coverage, \tname achieves
\covtfffvstt-\covtfffvspf higher coverage compared to the baselines across both \dl
libraries. Considering validity rate, \tname\ achieves
\torchDeltaRatioPathfinder{}x-\torchDeltaRatioAce{}x higher validity rate for
PyTorch and \tfDeltaRatioTT{}x-\tfDeltaRatioPathfinder{}x higher validity rate
for TensorFlow compared to the baselines.
Considering the ability to reveal crashes, we find that \tname{} finds more unique
crashes than any other baseline. On \torch{}, \tname{} discovers 5, 5, and 17
more crashes than \acetest{}, \pathfinder{}, and \ttfuzz{}, respectively. Only
\ttfuzz{} detects a single crash missed by \tname{}, while \acetest{} and
\pathfinder{} find none. On \tf{}, \tname{} discovers 18, 11, and 12 more
crashes than \acetest{}, \pathfinder{}, and \ttfuzz{}, respectively.
%
We also conduct long-running fuzzing campaigns with \tname\ (8 hours per API) on
the latest versions of \torch\ and \tf\ to find bugs on their APIs. During these
campaigns, \tname revealed a total of \numBugs{} new bugs;
\numBugsTorch\ in \torch\ and \numBugsTF\ in \tf. Of these, developers fixed \numBugsFix.


\subsection{Contributions}
This paper makes the following contributions:

\begin{itemize}[topsep=0ex,itemsep=0pt,leftmargin=1em]

\item[\Contrib{}]\textbf{Study.} We report the first in-depth study of the
  use of coverage-guided fuzzing (\cgf) to test \dl\ APIs. To enable the study
  on a large number of APIs, we developed \tname, an LLM-based approach to
  automatically synthesize test harnesses for \cgf. We compare \cgf\ against
  three \sota\ techniques using standard evaluation metrics. Results
  indicate that \cgf\ is a competitive technique for testing \dl\ APIs and
  researchers should use it as a comparison baseline when evaluating API-level
  fuzzers. Section~\ref{sec:lessons} elaborates other lessons from this study.

\item[\Contrib{}]\textbf{Bugs.} We report \numBugs{} new bugs in \torch\ and
\tf\ by applying \tname\ ``in the wild'', of which \numBugsFix\ have been fixed by
developers. 

\item[\Contrib{}]\textbf{Artifacts.} We make our artifacts -- including
  datasets and scripts -- publicly available: \RepoURL.

\end{itemize}

\noindent


\section{Background}
\label{sec:background}
\subsection{Tensor Representation}
\label{sec:tensor}

A tensor is a data structure for representing and manipulating machine
learning data in deep learning
frameworks~\cite{goodfellow2016deep}. Tensors generalize vectors
(first-order tensors) and matrices (second-order tensors) to arbitrary
dimensions, providing a unified mathematical framework for deep
learning computations \cite{kolda2022tensor,
  bengio2013representation}.
Tensors are characterized by four essential properties: (1)
\textbf{Data Type} determines both the memory representation and the
precision of computation. Common data types include floating-point
(e.g., \texttt{float32}), integer (e.g., \texttt{int64}), and boolean
values; (2)~\textbf{Number of Dimensions (nDim)} defines the number of
axes in the tensor; (3)~\textbf{Shape} specifies the size of each
dimension. For instance, a tensor with shape $(3, 4, 5)$ has 3
elements along its first dimension, 4 elements along its second
dimension, and 5 elements along its third dimension. In total, the
tensor contains 60~(=$3\times{}4\times{}5$) elements;
(4)~\textbf{Elements} constitute the individual values stored within
the tensor. The total number of elements is the product of all
dimension sizes.

\subsection{Deep Learning Library Organization}
\label{sec:deep-learning-lib-org}

\begin{wrapfigure}[8]{r}{0.25\textwidth}
  \vspace{-4ex}
  \centering
  \includegraphics[trim={0 15cm 28.25cm 0},clip,width=0.2\textwidth]{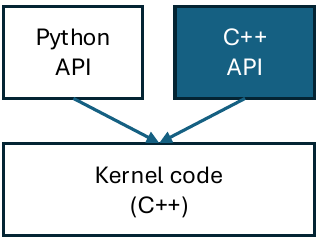}
  \caption{Layered organization of a Deep Learning Library.}
  \label{fig:dllib-org}
\end{wrapfigure}
Figure~\ref{fig:dllib-org} shows the typical layered organization of a
\dl\ library.  The kernel code --written in low-level languages like
C++-- implements the main functionalities of a library. A \dl\ library
provides different versions of a kernel to different devices, e.g.,
GPU kernel (for performance) and CPU kernel (for portability). Library
users access these kernels through high-level APIs, often in Python or
C++.  Python and corresponding C++ APIs, in principle, implement the
same ``glue'' functionality to access the kernel. This work tests the kernel code through C++ APIs,
highlighted at the top right of the figure.

\subsection{Coverage Guided Fuzzing}
\label{sec:coverage-guided-fuzzing}

Coverage-guided fuzzing (\cgf) is a testing technique that uses
coverage information to guide test input data generation~\cite{afl,
  libfuzzer, honggfuzz}. \cgf uses a queue to store input data, which
is typically encoded as a byte array. In a given iteration, \cgf
dequeues an element from the queue, generates mutations for that
input, translates the input byte array to the actual input data, and
executes the target program on those inputs.  The inputs that uncover
new branches are added back to the queue for further
processing. Fuzzing tools expect developers to instrument the program
to be tested to collect coverage information (e.g., LLVM compiler
option \texttt{-fsanitize=fuzzer-no-link}). Additionally, these tools
expect developers to provide a \emph{test harness} that (1) translates
a byte array representation to the representation the program uses;
(2) invokes the target program on that input; and (3) checks the
correctness of the corresponding outputs.

Fuzzing is a computationally-intensive task. To accelerate bug
finding, fuzzing tools often use in-process fuzzing~\cite{libfuzzer}
where the test harness and the target program are compiled together to
generate a single executable. 
(see LLVM \texttt{-fsanitize=fuzzer} compiler option). 
With that, the program can be exercised through
function calls (\ie{}, target program and fuzzer in the same process) as opposed to system calls (\ie{}, target program and fuzzer in different processes). We choose \libfuzzer{} as
our fuzzing engine. However, there is no fundamental reason precluding
the use of other engines, such as \afl~\cite{aflnew} and
HonggFuzz~\cite{honggfuzz}.




\section{Illustrative Example}
\label{sec:illustrative-example}
\begin{figure}[htbp]
  \centering
  \begin{minipage}[t]{0.48\textwidth}
    \begin{subfigure}[t]{\textwidth}
\begin{lstlisting}[language=C++, caption={Test harness template.}, label={lst:testharness-template}, numbers=left, numberstyle=\tiny, basicstyle=\ttfamily\tiny, breaklines=true, columns=fullflexible, showstringspaces=false, linewidth=\linewidth, aboveskip=0.25ex, belowskip=0.25ex, captionpos=b, frame=bottomline, framerule=0.3pt]
#define MAX_RANK 4
#define MIN_RANK 0

extern "C" int LLVMFuzzerTestOneInput(const uint8_t* data, size_t size) {
 size_t offset = 0;
 tensorflow::Scope root = tensorflow::Scope::NewRootScope();
 try {
   // --- 1. Create Fuzzed Input Tensors ---
   // --- 2. Call the Target TensorFlow API ---
   // --- 3. Run the TensorFlow API ---
   if (!status.ok()) {
     std::cout<<"TF Exception:"<<status.ToString()<<std::endl;
     return -1; } // will discard the seeds
 } catch (const std::exception& e) {
   tf_fuzzer_utils::logError("TF Exception: " + 
     std::string(e.what()), data, size);
   return -1;  
 } // end catch
 return 0; 
} // end try
\end{lstlisting}
    \end{subfigure}
  \end{minipage}%
  \hfill
  \begin{minipage}[t]{0.48\textwidth}
    \begin{subfigure}[t]{\textwidth}
\begin{lstlisting}[language=C++, caption={Some helper functions.}, label={lst:helper-function}, numbers=left, numberstyle=\tiny, basicstyle=\ttfamily\tiny, breaklines=true, columns=fullflexible, showstringspaces=false, linewidth=\linewidth, aboveskip=0.25ex, belowskip=0.25ex, captionpos=b, frame=bottomline, framerule=0.3pt]
tensorflow::DataType parseDataType(uint8_t selector) {
  tensorflow::DataType dtype;
  switch (selector % 20) {
    case 0:
      dtype = DT_FLOAT;
      break;
    // other 19 cases are omitted for brevity
  }
  return dtype; 
}
uint8_t parseRank(uint8_t byte) {
  constexpr uint8_t range = MAX_RANK - MIN_RANK + 1;
  uint8_t rank = byte % range + MIN_RANK;
  return rank; 
}

std::vector<int64_t> parseShape(const uint8_t* data, size_t& offset, size_t total_size, uint8_t rank) {...}

void createTensor(tensorflow::Tensor& tensor, const uint8_t* data, size_t& offset, size_t total_size) {...}
\end{lstlisting}
    \end{subfigure}
  \end{minipage}
  \vspace{-2ex}
  \caption{Some of the artifacts for prompt creation.}
  \label{fig:three-listings}
\end{figure}

This section briefly summarizes the process \tname\ uses to synthesize a test
harness (Section~\ref{sec:harness-generation}), a concrete example of a test
harness for the \torch API \CodeIn{fmod} (Section~\ref{sec:fmod-example}), and
how \cgf\ uses that harness to test APIs (Section~\ref{sec:cgf-with-harness}).

\subsection{Test Harness Generation with \tname}
\label{sec:harness-generation}

\sloppy
\tname\ uses different artifacts to generate test
harnesses that create a high number of semantically-valid inputs. First, \tname\
uses a C++ template that contains the overall structure of a test harness
(Listing~\ref{lst:testharness-template}), including setup code for each
framework, and checks for different exceptions. \Comment{consistency checks of
its outputs on different devices (e.g., CPU/GPU).} Second, \tname\ uses the API
signature and documentation to guide the generation of code that parses raw
input data and generates semantically-valid inputs. For example, for the API
\CodeIn{acos} the documentation states \emph{``Computes acos of parameter x
element-wise... Input range is [-1, 1]... x: A Tensor. Must be one of the
following types: bfloat16, ...''}~\cite{acostensorflow}. Third, \tname\ defines
and uses helper functions that parse raw data into objects of a certain
type~(Listing~\ref{lst:helper-function}).

\begin{figure*}[!htb]
  \centering
  \includegraphics[width=\textwidth]{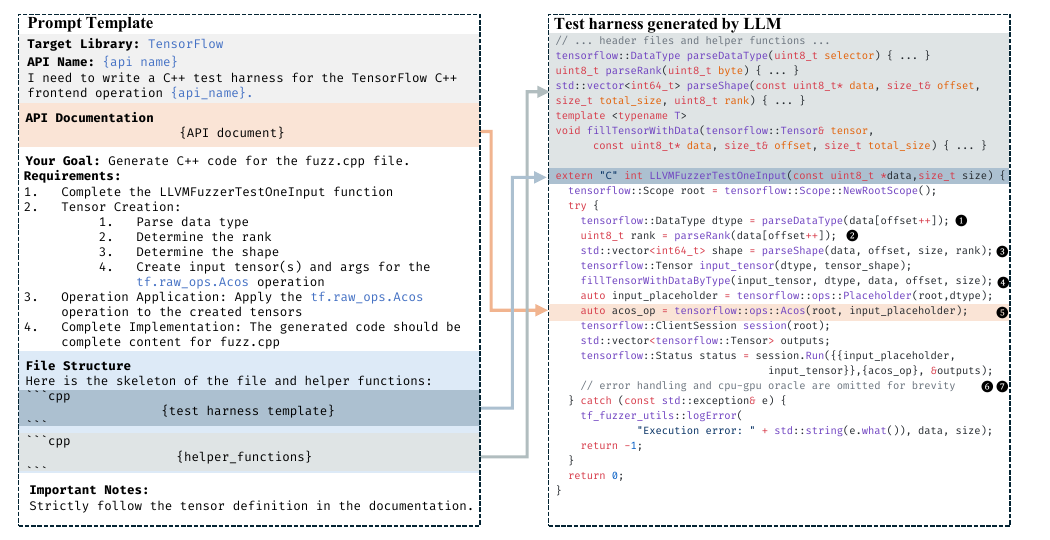}
  \caption{Prompt for generating a test harness for the API
    \CodeIn{tf.raw\_ops.Acos} and corresponding harness.}
  \label{fig:prompt}
\end{figure*}

Figure~\ref{fig:prompt} shows the various elements used in the creation of the
prompt we use to generate a test harness for the API
\CodeIn{tf.raw\_ops.Acos}~\cite{acostensorflow}. 
Figure~\ref{fig:prompt} (left) shows our prompt template for
generating test harnesses. It specifies the target library, API name, its
documentation, a template describing the required format for the test harness
(Listing~\ref{lst:testharness-template}), and a set of helper functions
(Listing~\ref{lst:helper-function}). It also provides a set of
\emph{requirements} that instruct the LLM on how to use the provided artifacts
to generate a correct test harness (\eg{}, the steps for tensor creation and
calling the target API).
The right-hand side of Figure~\ref{fig:prompt} shows a test harness for the API
\CodeIn{fmod} that \tname\ generates using the prompt template shown on the
left-hand side of the figure. This API takes only one input tensor as input but
the parsing method \tname\ employs generalizes to APIs with arbitrary number and
type of parameters. Note that the parameters in these libraries are restricted to
numeric types and container types (e.g., tensor, list, and tuple) of numeric
elements. 

The test harness from Figure~\ref{fig:prompt}~(right) follows several steps to test the API. It parses the byte array (steps 1-3), creates a tensor object (step 4), calls the API on the parsed
inputs (step 5) with framework specific constructs (e.g., \tf\ \CodeIn{Session}),
and checks for exceptions (steps 6-7).

\subsection{An example test harness for the  \torch API \CodeIn{fmod} and a bug
it reveals}
\label{sec:fmod-example}
%
We elaborate on the test harness for the \torch API
\CodeIn{torch.fmod}, which is similar to the harness for
\CodeIn{tf.raw\_ops.Acos} shown in Figure~\ref{fig:prompt}. The API
\CodeIn{fmod} computes the element-wise remainder of division. It takes as input
one tensor and either a scalar or another tensor. The output is a tensor with
the same shape as the input tensor. The API supports tensors with up to 12 data types (e.g., \CodeIn{torch.float32},
\CodeIn{torch.int64}, etc.). The API has two constraints: (1) if the second
parameter is a tensor, it must have the same shape as the first parameter; and
(2) if the second parameter is a scalar, it must be non-zero.
%
Fuzzing with this harness reveals a bug that we report to PyTorch developers,
who confirmed the bug~\cite{bug-fmod}. The bug is manifested when calling the
API with a specific large negative number; the function crashes raising
\textit{``Floating point exception (core dumped)''} error. 

\subsection{\cgf\ with a \tname\ Test Harnesses}
\label{sec:cgf-with-harness}

\begin{figure}[t!]
    \centering
    \includegraphics[trim={0.2cm 0.5cm 0.2cm 0 },clip,width=\textwidth]{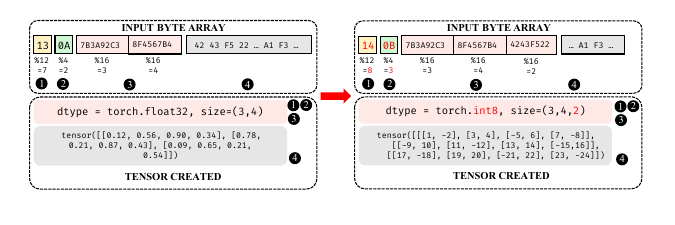}
    \caption{Input representation in test harness. Byte values in the input array are shown in hexadecimal.}
    \label{fig:illustrative-example-1-top}
\end{figure}

Now we present how the test harness for \CodeIn{fmod} works with a CGF fuzzer
and effectively generates semantically-valid inputs via mutation.
Figure~\ref{fig:illustrative-example-1-top} shows the layout of a byte array
representing an input for the API \CodeIn{torch.fmod} and its mutation
process~\cite{pytorch_fmod}. In step \circled{1}, \tname{} reads the first byte
of the byte array to determine the type of the tensor. The value \CodeIn{0x13}
maps to 7 via a modulo 12 operation (\tname\ supports 12 PyTorch data types) and
encodes the data type \CodeIn{torch.float32}. In step \circled{2}, \tname{}
extracts 2 as the number of dimensions in the tensor. It uses modulo 4 in that
case as that is the default setting of \tname\ for number of dimensions.
In step \circled{3}, \tname{} reads the shape of the tensor from two sequences
of four bytes. As before, \tname\ applies modulo 16 operations -- as that is
the bound on the maximum length of each dimension -- to find that the tensor has
a (3, 4) shape. Finally, in step \circled{4}, \tname{} extracts the 12
(=$3\times{}4$) \CodeIn{float32} values (4 bytes each) from the byte array and
arranges them according to the determined structure, creating a complete tensor
with floating-point values from the rest of the input byte array. 



It is worth noting that changes in the preamble of the input byte array
affect the tensor object to be created, as shown at the right-hand side of
Figure~\ref{fig:illustrative-example-1-top}. For example, let us
suppose the fuzzer mutates the first byte (\circled{1} in the input byte array)
from \CodeIn{0x13} to \CodeIn{0x14} resulting in a change from \CodeIn{float32}
to \CodeIn{int8}. In that case, \tname{} would consume 12 bytes (=12$\times$1)
instead of 48 bytes (=12$\times$4) to define the elements of the
$3\!\times{}\!4$ tensor. Consequently, the numeric elements of the generated
tensor would change. A more dramatic change occurs if the fuzzer changes the
second byte (\circled{2} in the input byte array) from \CodeIn{0x0A} to
\CodeIn{0x0B}. In that case, the shape would change to (3, 4, 2), affecting the
amount of data to be read from the input byte array. \tname\ uses a fixed input
byte array size of 128 bytes. When the required data exceeds available bytes,
missing values default to 1. Recall that \cgf\ fuzzers are capable of
identifying which of the generated inputs contribute to coverage and should be
retained (\textsection~\ref{sec:coverage-guided-fuzzing}). We additionally avoid retaining in the fuzzing queue inputs that result in runtime exceptions of the API as those are likely invalid.

\section{Method}
\label{sec:FlashFuzz}
\begin{figure*}[!tb]
  \centering
  \includegraphics[trim={0 0.6cm 0cm 0},clip,width=\textwidth]{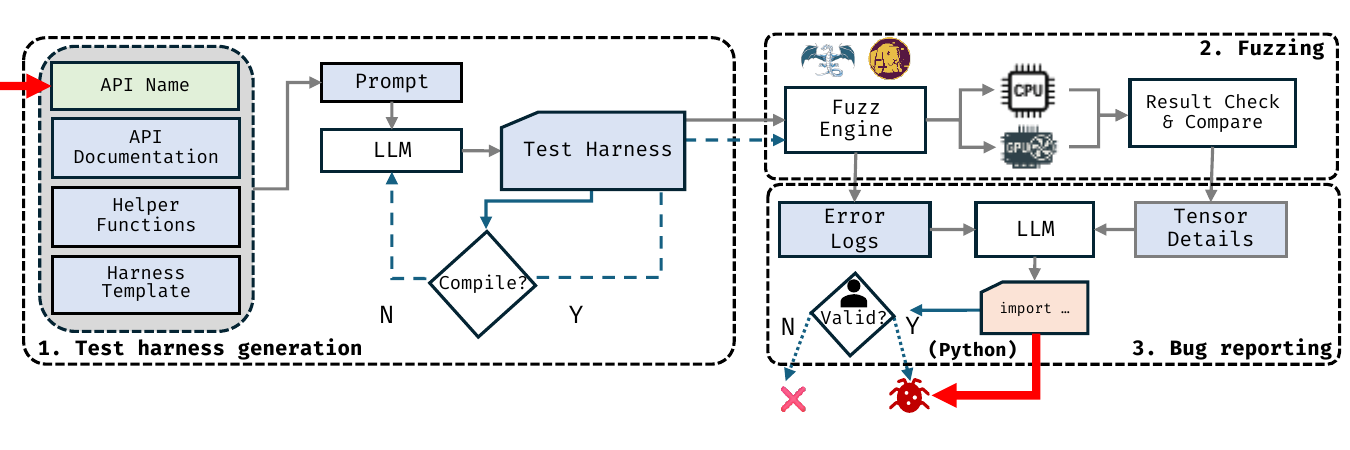}
  \caption{Workflow of Method Used to Fuzz Deep Learning Library APIs.}
  \label{fig:flashfuzz}  
\end{figure*}

This section describes the method we use to conduct the study from this paper.
Figure~\ref{fig:flashfuzz} shows the workflow we follow to test \dl\ library APIs.
First, the \emph{test harness generator}~(\textsection~\ref{tech:synthesis})
produces a test harness for a given API. Second, we start a \emph{fuzzing
campaign}~(\textsection~\ref{tech:fuzzing}) using the test harness obtained in
the previous step.
The discrepancies detected in this stage can be false alarms. Hence, in the
third step, we verify if the crash can be observed through the execution of the
C++ API and the corresponding Python API. We elaborate these steps next.

\subsection{Test Harness Generator}
\label{tech:synthesis}

We develop a test harness generation method to support this study. The method is
based on three observations: (1) the majority of the functionality of a \dl\ API
is in the kernel code, which is implemented in C++~\cite{PathFinder_ICSE25}
(\textsection~\ref{sec:deep-learning-lib-org}); (2) \dl\ APIs typically manipulate
tensor-like data structures; and (3) recent experiments have shown that Large
Language Models are capable of generating test harnesses capable of parsing raw
data into structured
inputs~\cite{ai-powered-bug-huntint,vikram2024largelanguagemodelswrite,lyu2024prompt}.

Generating semantically-valid inputs to test \dl\ library APIs can be
challenging. First, the test harness must be capable of converting array of
bytes (\ie{}, the input of highly-optimized \cgf{}s) into the data structures
that an API uses (\ie, tensor-like data structures). Second, the generated
inputs should satisfy complex input constraints of an API. To mitigate these
problems, \tname\ prompts an LLM with the API metadata, a C++ harness template,
and helper functions, which restricts the space of possible harnesses the LLM
can generate. 


The leftmost box from Figure~\ref{fig:flashfuzz} illustrates the LLM-based
approach that \tname\ employs to synthesize test harnesses. First, the user
provides an API name as input. Second, \tname\ uses the name of the API, its
documentation, helper functions (e.g., the function \CodeIn{parseTensorData} parses a tensor object from a
byte array), and a template for the harness to formulate a prompt describing the
task for the LLM. Third, \tname\ uses the prompt to query the LLM. Finally,
\tname\ checks if the obtained harness is plausible by (1) checking if it
compiles and (2) checking if the target API is the focal method in the harness.
If one of these criteria is not met, the prompt is updated with a description of
the observed problem. This process repeats until succeeding or reaching a budget
on the number of iterations. If successful, \tname\ reports a harness on output.



\MyItPara{API documentation}
\tname\ expects the \dl\ library to provide API documentation through the
attribute \CodeIn{\_\_doc\_\_}. For example, \tname\ retrieves the documentation
of the arc cosine API by reading the field
\CodeIn{tf.raw\_ops.Acos.\_\_doc\_\_}.
This documentation (\textsection~\ref{sec:harness-generation}) provides valuable information about the input constraints for
the API. 



\MyItPara{Helper functions}
Helper functions provide the infrastructure for converting raw binary input data
into structured program data. They enable creation of tensors with varying data
types, dimensional ranks, shapes, and values. The main helper functions we
develop are:
\begin{packed_enumerate}
  \item \CodeIn{parseDataType}: maps a byte to a scalar data
    type;
  \item \CodeIn{parseRank} extracts the number
    of dimensions of a tensor;
  \item \CodeIn{parseShape} extracts the size of each dimension of a
    tensor;
  \item \CodeIn{createTensor} uses information obtained with other
    helper functions to create a tensor object.
\end{packed_enumerate}

We provide these methods as a set of utility functions that the LLM can use. It
is important to note that the LLM can modify the definition of a helper
function. In the arc cosine example from Figure~\ref{fig:prompt}, the LLM uses
the API documentation to modify the helper function \CodeIn{parseDataType}
making it wrap the numeric parameter \CodeIn{select} within the 0-5 range and
return a \tf\ type that is among the options listed in the documentation (e.g.,
\CodeIn{bfloat16}, \CodeIn{half}, etc.).



\MyItPara{Harness template}
The harness template defines the skeleton for the test harness that
the LLM produces on output. It incorporates necessary header files to
compile the harness, namespace declarations, and standardized elements
to give structure to the test harness the LLM generates.

\begin{figure}[htbp]
  \centering
  \includegraphics[trim={0 12.5cm 4.6cm 0},clip,width=\columnwidth]{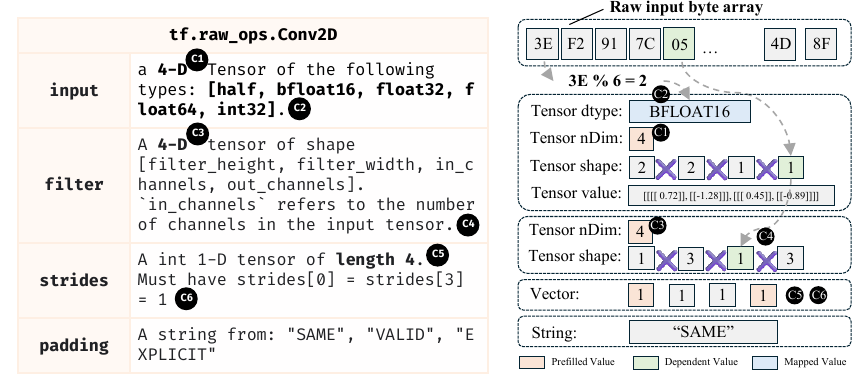}
  \caption{Illustration of LLM transformation patterns.}
  \label{fig:llm-fix}
\end{figure}

\subsubsection{Transforming Fuzzer-Generated Input Bytes}.~We observe that the LLM employs
three patterns to transform fuzzer-generated input byte arrays for a certain
API. Figure~\ref{fig:llm-fix} illustrates the three patterns on an input for the
API \CodeIn{tf.raw\_ops.Conv2D}, which has complex constraints~\cite{conv2d}.
This API requires multiple input arguments including an input tensor
(\CodeIn{input}), a filter tensor (\CodeIn{filter}), and a strides vector
(\CodeIn{strides}). The main constraints of the API are: (\textbf{c1}) the input
tensor must be 4-dimensional, (\textbf{c2}) the data type of the input tensor
must be one among six options, (\textbf{c3}) the filter
tensor must be 4-dimensional, (\textbf{c4})  the filter tensor must have a shape
compatible with the input tensor. (\textbf{c5}) the strides vector must contain
exactly four elements, (\textbf{c6}) both the first and last elements of the
strides vector must be 1.


\begin{packed_enumerate}
\item \textbf{Value Range Mapping.} The raw data may contain arbitrary numerical
values outside the range admitted by a parameter of an API. The LLM-generated
test harness addresses this issue by mapping out-of-range values to in-range
values of a numerical API parameter. Figure~\ref{fig:llm-fix} highlights the
usage of the modulo operator to enforce constraint \textbf{c2}. The operator
maps the first byte in the input array to one of the six choices of data types
listed in the documentation of \CodeIn{Conv2D} (corresponding to elements in the
\CodeIn{tensor} parameter of the API). We observe this kind of transformation in the helper function \CodeIn{parseDataType}.


\item \textbf{Value Prefilling.} Certain APIs require specific values as
input parameters. The constraints \textbf{c1}, \textbf{c3}, and \textbf{c6}
from \CodeIn{Conv2D} illustrate that. The LLM-generated harness
automatically identifies these constraints and prefills the necessary values
in the input arguments to ensure proper API functionality. For example,
Figure~\ref{fig:llm-fix} shows that the vector \CodeIn{strides} is prefilled
with the value one at both ends (constraint \textbf{c6}). Likewise, the
figure also shows that the attributes \CodeIn{nDim} from the tensors
\CodeIn{input} and \CodeIn{filter} are initialized with four dimensions
(\textbf{c1} and \textbf{c3}).

\item \textbf{Value Dependence.} The LLM-generated harness
demonstrates the capability to inherit appropriate values from the
surrounding context, ensuring that dependent parameters maintain
consistency with their parent operations or previously established
values. In Figure~\ref{fig:llm-fix}, the shape of the
\CodeIn{filter} tensor is derived from the channel dimension of
the \CodeIn{input} tensor, \CodeIn{in\_channels}.  More
specifically, \CodeIn{Conv2D} requires the size of the last
dimension of the \CodeIn{input} tensor (\CodeIn{in\_channels},
which equals to 1) to match the size of the third dimension of the
\CodeIn{filter} tensor, which is also 1. The \CodeIn{in\_channels}
represents the number of channels that the convolution
processes. The generated harness reads the input tensor's
channel count and automatically sets the filter's corresponding
dimension, satisfying constraint \textbf{c4} instead of using
values from the raw data.
\end{packed_enumerate}


\subsection{Fuzzing}
\label{tech:fuzzing}

\tname\ uses standard coverage-guided fuzzing
(\textsection~\ref{sec:coverage-guided-fuzzing}) to generate
inputs. As in prior work (\textsection~\ref{sec:related}),
\tname\ uses differential testing oracles to detect bugs across
different devices (i.e., CPU and GPU) during fuzzing. Those oracles
observe two kinds of divergences: (1)~\textbf{Output.}~\tname\ flags a
potential bug when outputs differ above a threshold of relative
tolerance $10^{-2}$ and absolute tolerance $10^{-3}$. This value is
consistent with those used in previous deep learning library testing
research~\cite{Deng_ETAL_ISSTA23, Wei2022};
(2)~\textbf{Exception.}~\tname\ flags a potential bug when the
execution on a device raises an exception while the other proceeds
without errors.


\subsection{Bug Reporting}
\label{tech:validation}

The preceding steps aim to find inputs that cause a C++ API to exhibit unexpected behavior. Unfortunately, \emph{false alarms} can occur. Despite our efforts to generate test harnesses that avoid invalid inputs, several sources of imprecision remain, including incomplete documentation and the inherent imprecision of LLM-generated harnesses. When available, \tname\ leverages the input-validity checkers provided by \dl libraries' C++ APIs (e.g., \torch). These checkers raise runtime errors that \tname\ can detect by parsing log messages and classifying the run as an ``invalid input'' execution. Our test-harness handles such errors. However, this design is not uniformly adopted across \dl\ libraries; \tf, for instance, performs portions of its validity checking only in the Python APIs. Consequently, some inputs that reach this stage may still be spurious when using \tname\ with \tf.

To mitigate this issue, \tname\ reports a bug only if it can be reproduced via the Python APIs. All bugs reported by \tname\ are manually triaged through a three-step process. First, we reproduce the crash by re-executing the test harness with the crashing input and verifying the error messages and stack traces in the logs. Second, we manually instrument the test harness with additional debug output. Third, we rerun the instrumented harness to collect the necessary diagnostics and attempt to reproduce the bug through the Python API, aided by an LLM chatbot. Finally, we prepare a bug report to submit to developers including the Python script to reproduce the bug and a description to explain the problem.



\section{Evaluation}
\label{sec:evaluation}

\noindent{}We answer the following research questions:

\DefMacro{eq-coverage}{How does \tname\ compare against the \sota\
on standard evaluation metrics?}

\DefMacro{eq-ablation}{How important are the different components of \tname{}?}

\DefMacro{eq-in-the-wild}{How effective is \tname\ in
  revealing new bugs in the latest releases of popular \dl\ libraries?}



\noindent\textbf{\EQ{1}:~}\UseMacro{eq-coverage}\\
\noindent\textbf{\EQ{2}:~}\UseMacro{eq-in-the-wild} \\
\noindent\textbf{\EQ{3}:~}\UseMacro{eq-ablation}\\


\EQ{1} evaluates how \tname\ compares against \sota\ techniques on standard
metrics from the literature (in API-level fuzzing of \dl libraries), namely
coverage, validity ratio, and the ability to reveal bugs.
%
%
%
\EQ{2} reports on \tname's ability to uncover bugs in the latest versions of
\torch\ and \tf. 
\EQ{3} reports an ablation study to assess the impact
of \tname's design choices.
%


\subsection{Answering \EQ{1}:~\UseMacro{eq-coverage}}

    
    

    

    


We compare \tname\ and \sota\ techniques using standard
evaluation metrics from the literature.

\MyPara{Comparison Baselines} We compare \tname\ against three recently-proposed
fuzzing techniques for finding bugs in deep learning libraries, namely
\emph{\acetest}~\cite{Shi_2023}, \emph{\pathfinder}~\cite{PathFinder_ICSE25},
and \emph{\ttfuzz}~\cite{Deng_ETAL_ISSTA23}. We select \acetest\ and
\pathfinder\ because they analyze the kernel implementation of the operators in
the backend of the libraries, like \tname\ does
and we select \ttfuzz\ because it is a well-known representative of a technique
that uses Large Language Models (LLMs) to generate
inputs.\footnote{FuzzGPT~\cite{deng2024large} claims to outperform \ttfuzz\ but
the tool is not publicly available} It is worth noting that \ttfuzz\ was
initially configured with \CodeIn{facebook/incoder-1B}, which is relatively
outdated (released April 2022). So, for a fair comparison, we communicated with
the authors of \ttfuzz\ and updated \ttfuzz\ LLM to use
\CodeIn{qwen2.5-coder:7b}~\cite{hui2024qwen25codertechnicalreport}, which is
newer (release July 2025) and larger compared to the original model. We
validated that \ttfuzz\ configured with the qwen model outperforms the original
configuration in terms of coverage, validity ratio and the ability to find bugs.
We reported the results to the authors.
To avoid bias, we use the scripts from the artifacts of the comparison baselines
to run their experiments and obtain measurements. We release the corresponding
patches as public forks of \acetest{}\footnote{\url{https://github.com/ncsu-swat/ACETest}}
and \ttfuzz{}.\footnote{\url{https://github.com/ncsu-swat/TitanFuzz}}

\MyPara{APIs Analyzed} Different techniques support different sets of APIs.
Table~\ref{tab:num_target_apis} shows the breakdown of APIs that \tname\
currently supports and the number of APIs in common with the comparison
baselines. 
To contextualize the scope of our targets, we quantified how many backend APIs
are eligible for \tname\ and how many are currently supported. For PyTorch, we
programmatically enumerated backend-facing APIs (e.g., \CodeIn{torch.Tensor.*},
\CodeIn{torch.nn.*}, \CodeIn{torch.fft.*}, \CodeIn{torch.linalg.*},
\CodeIn{torch.special.*}, \CodeIn{torch.jit.*}, and storage classes) and
obtained a total of 1,576 APIs. Of these, \tname\ currently supports \rqOneTorchTotalAPIs. For
TensorFlow, we enumerated all operators under \CodeIn{tf.\_api.v2.raw\_ops} and
obtained a total of 1,452 "raw operators". Of these, \tname\ currently
supports \rqOneTFTotalAPIs. These totals reflect the set of APIs that \tname\ considers valid
targets for backend fuzzing. It is worth noting that we can construct a mapping
between frontend and backend APIs by leveraging the consistent naming
conventions adopted by libraries.
\begin{table}[h!]
  \small
  \centering
 \setlength{\tabcolsep}{7pt}  
 \caption{Number of APIs \tname supports (column ``\tname'') and number of APIs in
  common between \tname\ and each comparison baseline (columns ``$\cap$ <baseline-name>'').}
\label{tab:num_target_apis}
\begin{tabular}{lrrrr}
\toprule
& \tname & $\cap$ \acetest & $\cap$ \pathfinder & $\cap$ \ttfuzz  \\
\midrule
\torch & \rqOneTorchTotalAPIs & \rqOneTorchIntersectACE  & \rqOneTorchIntersectPathfinder & \rqOneTorchIntersectTTFuzz \\
\tf  & \rqOneTFTotalAPIs & \rqOneTFIntersectACE & \rqOneTFIntersectPathfinder & \rqOneTFIntersectTTFuzz \\
\bottomrule
\end{tabular}

\end{table}


\MyPara{Metrics} We use \emph{coverage}, \emph{validity ratio}, and \emph{bug
detection ability} as comparison metrics as they have been previously used in
the literature to evaluate API-level
fuzzers~\cite{Shi_2023,PathFinder_ICSE25,Deng_ETAL_ISSTA23}. For \emph{coverage} and \emph{validity ratio}, we run all the techniques on \torch~\torchvereval\ and \tf~\tfvereval{}. For \emph{bug detection ability}, we run the techniques on the latest library version, namely \torch~\torchverwild\ and \tf~\tfverwild{}.

\MyItPara{Coverage} As in the \pathfinder\ evaluation~\cite{PathFinder_ICSE25},
we measure branch coverage of C++ kernel code. To compute coverage, we
instrument the \torch and \tf libraries to report branch coverage by using the
LLVM flags \CodeIn{-fprofile\!-\!instr\!-\!generate} and
\CodeIn{-fcoverage\!-\!mapping} flags. We use llvm-cov~\cite{llvm-cov} to
analyze the coverage data and filter coverage information to focus only on
kernel functions: \CodeIn{aten/src/ATen/native} for \torch and
\CodeIn{tensorflow/core/kernels} for \tf. Filtering has been used in the
evaluation of prior work~\cite{PathFinder_ICSE25}; the rationale is to prevent
irrelevant, possibly large, portions of code from distorting coverage numbers.
We open-sourced a coverage analysis tool to facilitate future research on
coverage analysis for \dl\ libraries.\footnote{\url{https://github.com/ncsu-swat/Universal-DLL-Coverage-Collector}}
To obtain coverage data, we configured the techniques as follows. For \acetest
and \ttfuzz, we calculate coverage based on their generated files. For
\pathfinder, we modified their artifacts to use \CodeIn{llvm\!-\!cov} (instead of \CodeIn{lcov}~\cite{lcov}) to
ensure consistency in coverage measurement across all techniques.

\MyItPara{Validity ratio} As commonly done in the
literature~\cite{Shi_2023,PathFinder_ICSE25,Deng_ETAL_ISSTA23} due to the lack
of formal API specifications, validity ratio is defined as the percentage of
generated inputs that do not trigger input validation errors when executed in
the target API.

\MyItPara{Bug detection ability} To assess the ability of techniques to
reveal bugs, we count the number of unique crashes each technique generates,
i.e., inputs that cause the program to crash. We manually examine the stack
traces and error messages to deduplicate the crashes. 


\vspace{0.5ex}\noindent\textbf{Evaluation setup.}
For the generation stage of \ttfuzz, we used two machines equipped with AMD EPYC
7742 CPU and 8 NVIDIA RTX V100 GPUs. For the coverage calculation stage of
\ttfuzz and all the other experiments, we used two CPU-only machines equipped
with AMD EPYC 9684X CPU. \tname\ uses Claude Sonnet
4.0~\cite{anthropic2025claude4} for test harness generation.
\Comment{; we find that model
to perform well on that task.} We use a 10-minute fuzzing budget per target API.



\subsubsection{Results for coverage}
\begin{figure}[t!]
\centering
\captionsetup[subfigure]{justification=centering,singlelinecheck=false,font=small}
    \begin{subfigure}{\columnwidth}
       \centering
       \includegraphics[width=0.33\linewidth]{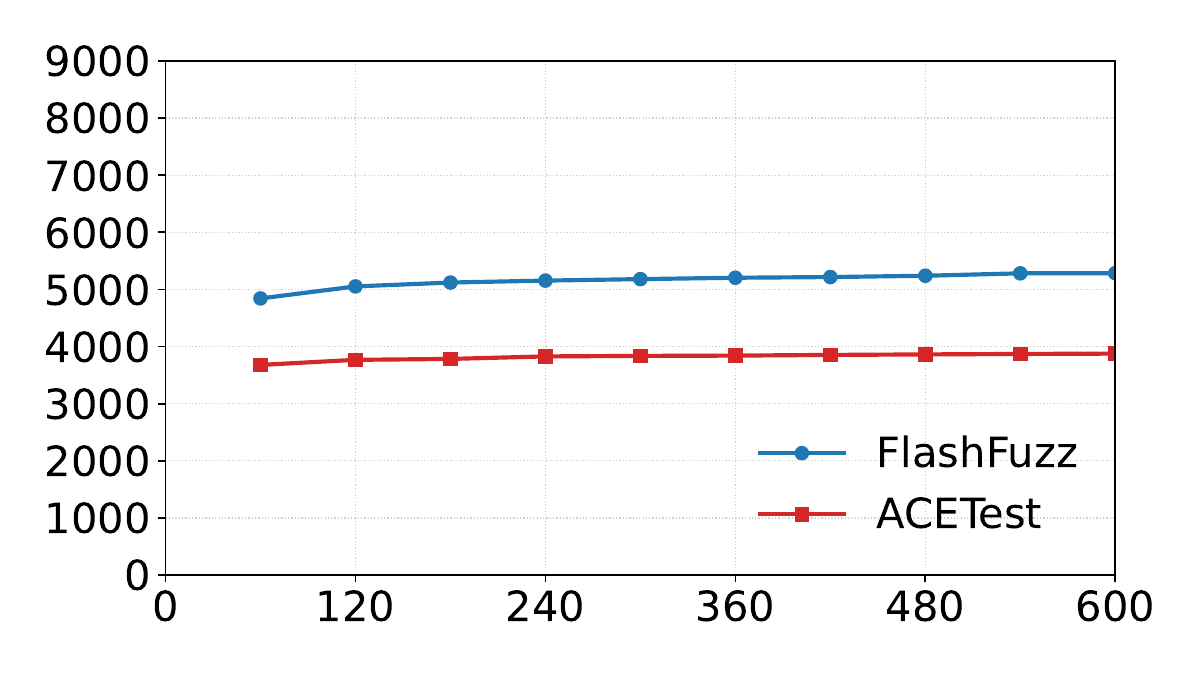}%
       \includegraphics[width=0.33\linewidth]{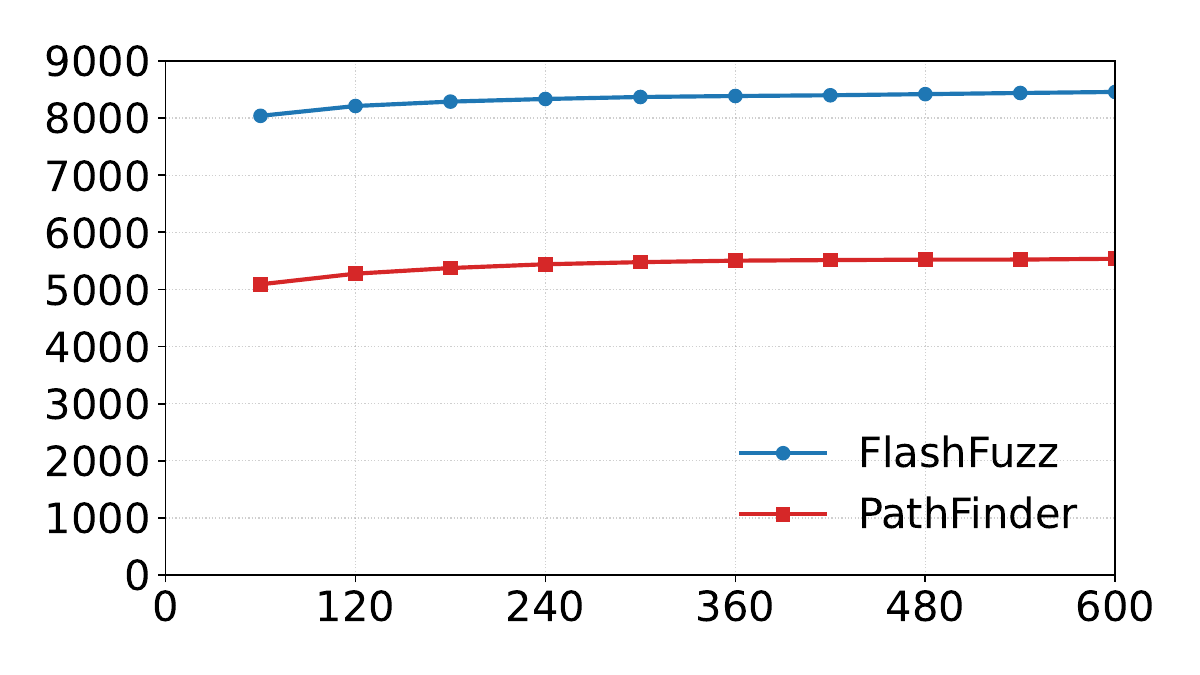}%
       \includegraphics[width=0.33\linewidth]{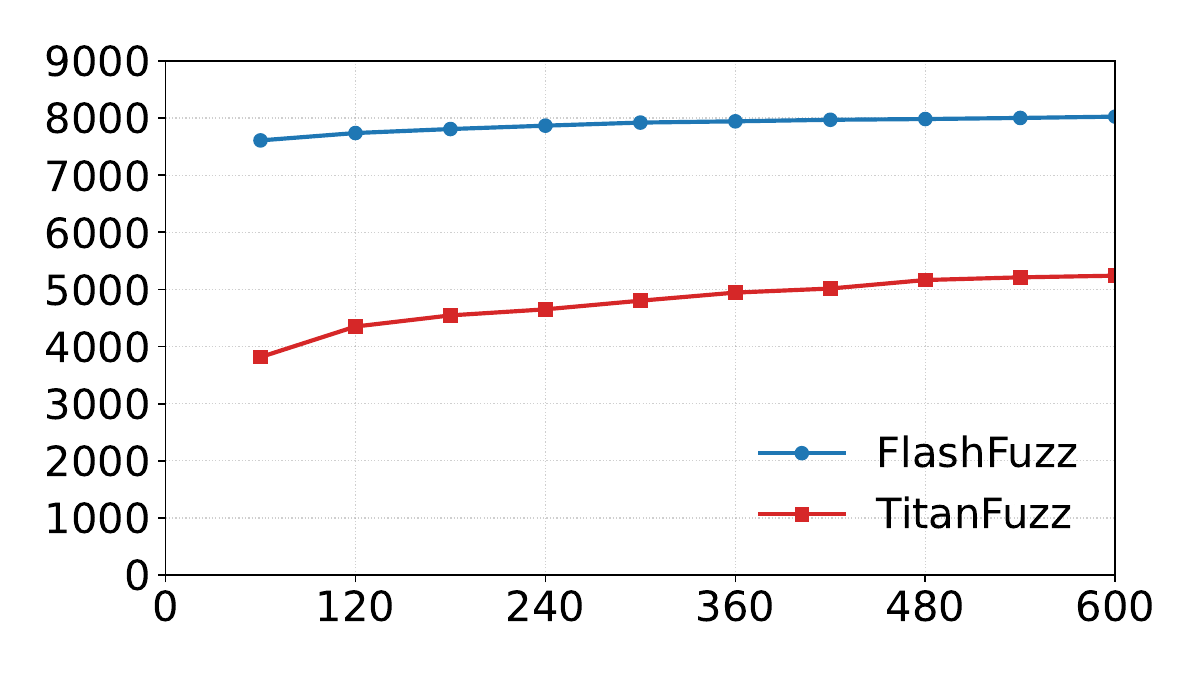}%
       \vspace{-0.75ex}
       \caption{\tname vs. \{\acetest{}, \pathfinder, \ttfuzz\} on \torch.\label{fig:rq1-cov-torch}}
    \end{subfigure}

    \begin{subfigure}{\columnwidth}
       \centering
       \includegraphics[width=0.33\linewidth]{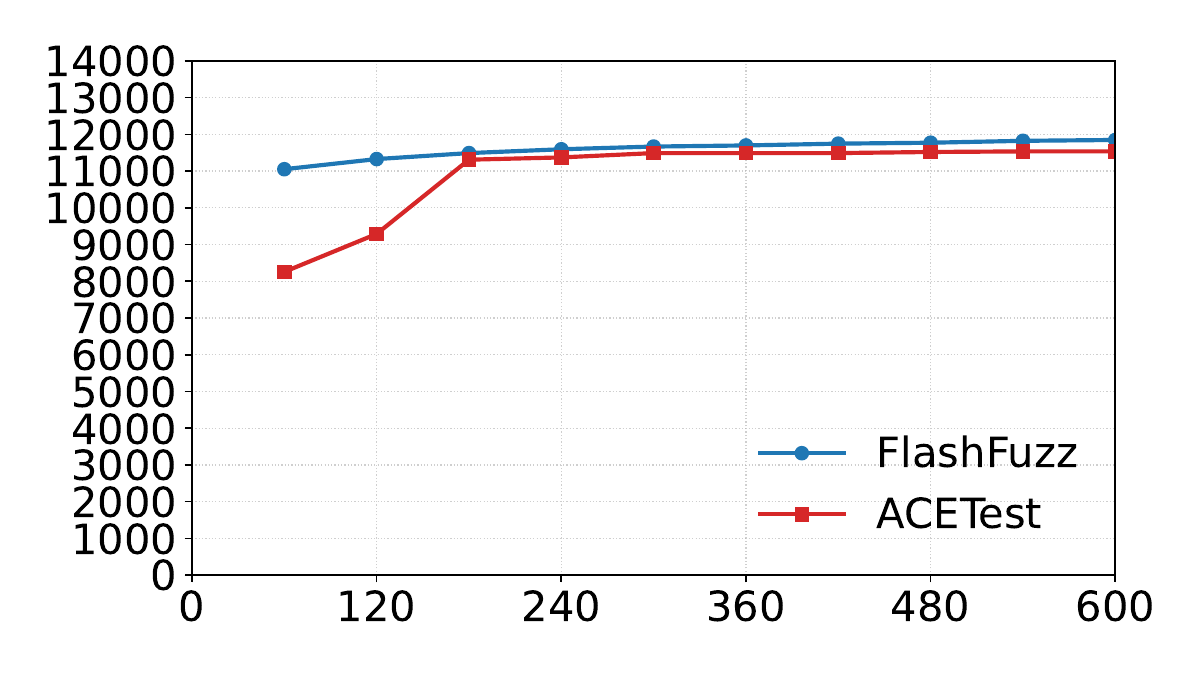}%
       \includegraphics[width=0.33\linewidth]{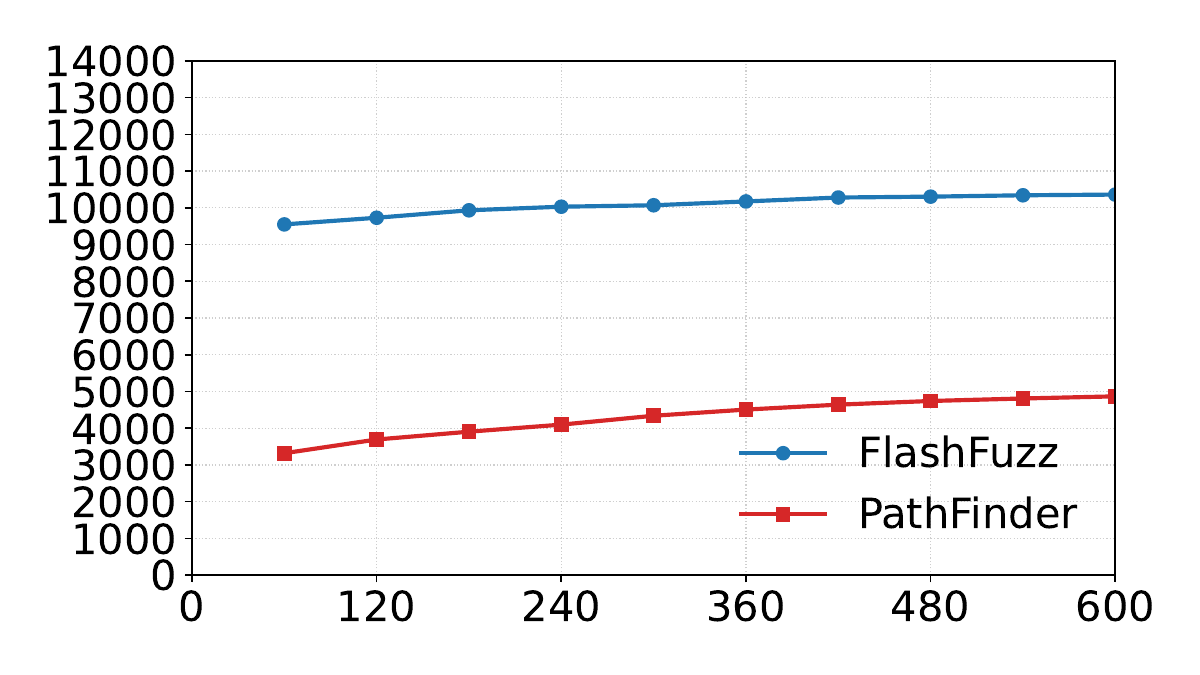}%
       \includegraphics[width=0.33\linewidth]{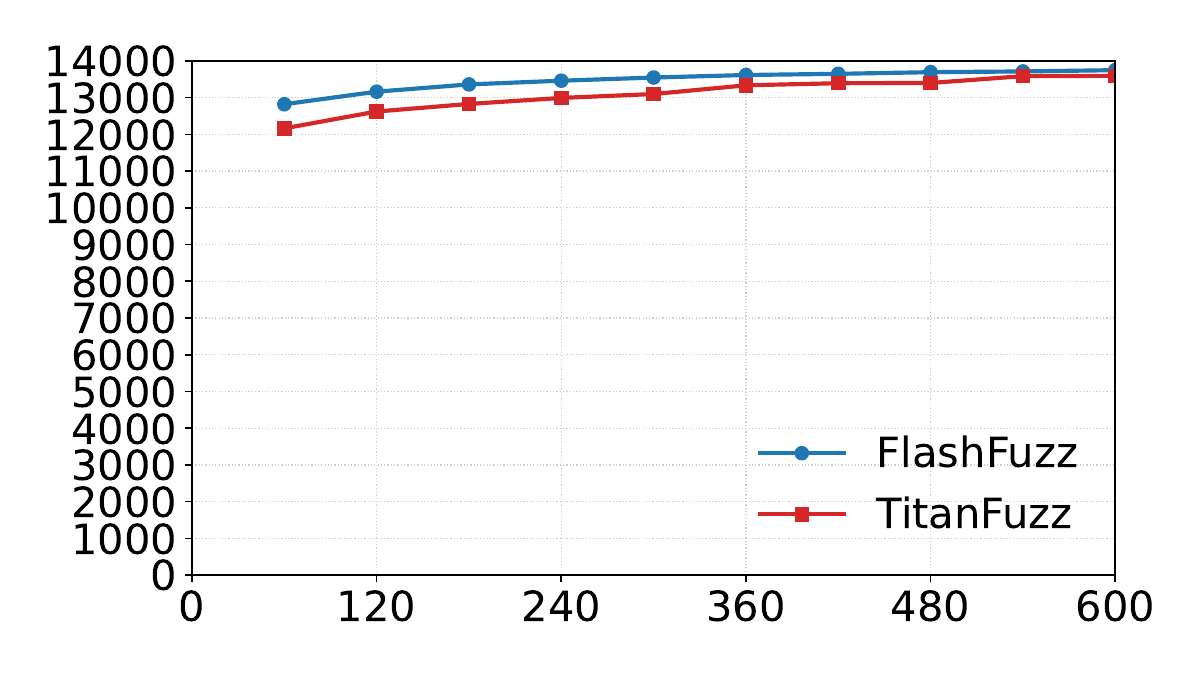}%
       \vspace{-0.75ex}
       \caption{\tname vs. \{\acetest{}, \pathfinder, \ttfuzz\} on \tf.\label{fig:rq1-cov-tf}}
    \end{subfigure}    
\caption{Coverage progress for \tname\ and comparison technique on
  \torch\ (top) and \tf\ (bottom).\label{fig:rq1-torch}\label{fig:rq1-tf}\label{fig:rq1-cov}}
\end{figure}

Figure~\ref{fig:rq1-cov} presents the progress of branch coverage for each
technique over time, with a budget of 10 minutes. We report results for each
pair of techniques on \torch\ and \tf. Recall that techniques support different
sets of APIs (Table~\ref{tab:num_target_apis}). To ensure a fair comparison, we
only consider the APIs that both techniques support. Each data point represents
the merged coverage over all APIs that both techniques support. Overall, results
show that \tname\ consistently outperforms all baselines in both \torch\ and
\tf. Note that the differences between \tname\ and \acetest\ on \tf\ are marginal
after 3 minutes (=180 seconds). Likewise, the difference between \tname\ and
\ttfuzz\ on \tf\ is marginal during the entire campaign.


\begin{table*}[!tb]
\centering
\caption{Total number of inputs each tool generates and their valid input rates.}
\label{tab:dl_tools_comparison}
\scriptsize
\tabcolsep=9pt
\begin{tabular}{l|r|r||r|r||r|r}
\toprule
\multicolumn{7}{c}{\textbf{\torch}} \\
\midrule
\textbf{Tool} & \textbf{\acetest{}} & \textbf{\tname{}} & \textbf{\pathfinder{}} & \textbf{\tname{}} & \textbf{\ttfuzz{}} & \textbf{\tname{}} \\
\midrule
\textbf{Total} & 9,267.4K & \specialcellr{158,027.2K\\\deltax{\torchDeltaTotalAce}} & 58,333.2K & \specialcellr{270,083.6K\\\deltax{\torchDeltaTotalPathfinder}} & 206.8K & \specialcellr{244,433.4K\\\deltax{\torchDeltaTotalTT}} \\
\midrule
\textbf{Valid} & 4,918.4K & \cellcolor{green}\specialcellr{104,971.4K\\\deltax{\torchDeltaValidAce}} & 36,126.7K & \cellcolor{green}\specialcellr{170,117.1K\\\deltax{\torchDeltaValidPathfinder}} & 126.6K & \cellcolor{green}\specialcellr{160,943.9K\\\deltax{\torchDeltaValidTT}} \\
\midrule
\textbf{Ratio (\%)} & 53.07 & \specialcellr{66.43\\\deltax{\torchDeltaRatioAce}} & 61.93 & \specialcellr{62.99\\\deltax{\torchDeltaRatioPathfinder}} & 61.22 & \specialcellr{65.84\\\deltax{\torchDeltaRatioTT}} \\
\midrule
\end{tabular}
\begin{tabular}{l|r|r||r|r||r|r}
\toprule
\multicolumn{7}{c}{\textbf{\tf}} \\
\midrule
\textbf{Tool} & \textbf{\acetest{}} & \textbf{\tname{}} & \textbf{\pathfinder{}} & \textbf{\tname{}} & \textbf{\ttfuzz{}} & \textbf{\tname{}} \\
\midrule
\textbf{Total} & 38,451.8K & \specialcellr{39,047.2K\\\deltax{\tfDeltaTotalAce}} & 20,392.2K & \specialcellr{34,662.0K\\\deltax{\tfDeltaTotalPathfinder}} & 110.7K & \specialcellr{59,252.4K\\\deltax{\tfDeltaTotalTT}} \\
\midrule
\textbf{Valid} & 14,843.5K & \cellcolor{green}\specialcellr{37,183.2K\\\deltax{\tfDeltaValidAce}} & 3,605.1K & \cellcolor{green}\specialcellr{32,866.4K\\\deltax{\tfDeltaValidPathfinder}} & 60.0K & \cellcolor{green}\specialcellr{54,549.7K\\\deltax{\tfDeltaValidTT}} \\
\midrule
\textbf{Ratio (\%)} & 38.60 & \specialcellr{95.23\\\deltax{\tfDeltaRatioAce}} & 17.68 & \specialcellr{94.82\\\deltax{\tfDeltaRatioPathfinder}} & 54.23 & \specialcellr{92.06\\\deltax{\tfDeltaRatioTT}} \\
\bottomrule
\end{tabular}
\end{table*}

\subsubsection{Results for validity ratio}
Table~\ref{tab:dl_tools_comparison} presents the cumulative results across all
APIs supported by a given pair of techniques, showing (1) the total number of
inputs each tool generates, (2) the number of valid inputs, and (3) the
percentage of valid inputs generated. We observe that the number of valid inputs
\tname\ generates within the time budget is at least \tfDeltaValidAce{} times higher
and at most \torchDeltaValidTT{} times higher compared to the baselines,
demonstrating \tname's efficiency. See highlighted cells.

In \torch, we observe that \tname produces valid inputs at similar ratios
compared to the baselines, but the number of valid inputs is significantly
higher in \tname. In \tf, the ratio of valid inputs of \tname\ is
\tfDeltaRatioTT{}x to \tfDeltaRatioAce{}x higher than the baselines. 

\subsubsection{Results for bug detection ability}

We compare \tname{} against each baseline by measuring the number of unique crash-triggering inputs. \tname{} discovers substantially more unique crashes than all baselines. On \torch{}, it uncovers 5 crashes compared to \acetest{} (0), \pathfinder{} (0), and 17 crashes compared to \ttfuzz{} (1), with no overlaps between \tname{} and any baseline. On \tf{}, it uncovers 21 crashes vs. \acetest{} (3), 16 vs. \pathfinder{} (5), and 25 vs. \ttfuzz{} (13), with overlaps of 2, 4, and 1 crashes, respectively.


\begin{mdframed}[style=mpdframe,frametitle=\UseMacro{eq-coverage}] \tname{}
outperforms all techniques on all metrics. Most importantly, the ability to detect bugs is significantly higher in \tname\ compared to the baselines.
\end{mdframed}


\subsection{Answering \EQ{2}:~\UseMacro{eq-in-the-wild}}








\begin{table}[t!]
  \caption{\label{tab:bug_stats}Bug statistics.}
  \vspace{-2ex}
  \footnotesize
  \begin{subtable}{\linewidth}
    \centering
    \caption{\label{tab:bug_status}Status of bugs reported to developers.}
    \vspace{-1ex}
    \begin{tabular}{ccccccc}
      \toprule
      \textbf{Framework} & \textbf{Reported} & \textbf{Duplicate}
      & \textbf{Rejected} & \textbf{Confirmed} & \textbf{Fixed} & \textbf{Pending} \\
      \midrule
      \torch  & 25  & 1  & 4  & \cellcolor{lightgray}20  & 7 & 0 \\
      \tf\    & 22  & 0  & 0  & \cellcolor{lightgray}21  & 1 & 1 \\
      \bottomrule
    \end{tabular}
    \vspace{1em}
  \end{subtable}
  
  \begin{subtable}{\linewidth}
    \setlength{\tabcolsep}{2pt} 
    \centering
    \caption{\label{tab:bug_types}Types of bugs (Confirmed + Pending).}
    \vspace{-1ex}
    \begin{tabular}{lccccccr}\toprule
      \multirow{3}{*}{\textbf{Framework}} &\multicolumn{6}{c}{\textbf{Bug Type}} &\multirow{3}{*}{\textbf{$\Sigma$}} \\ \cmidrule{2-7}
      &\multirow{2}{*}{\textbf{Aborted}} &\multirow{2}{*}{\textbf{Segfault}} &\multirow{2}{*}{\textbf{FPE}} &\textbf{Memory} &\textbf{Internal} &\textbf{Inconsistency Validity} & \\
      & & & &\textbf{Overflow} &\textbf{Exception} & \textbf{(CPU-GPU)} & \\\midrule
      \torch &1 &2 &5 &1 &0 &11 &20 \\
      \tf    &9 &6 &2 &0 &2 &3 &22 \\ \midrule
      $\Sigma$ &10 &8 &7 &1 &2 &14 &\cellcolor{lightgray}42 \\
      \bottomrule
    \end{tabular}

  \end{subtable}

\end{table}

Table~\ref{tab:bug_stats} shows the status of the bugs we reported to developers when fuzzing APIs ``in the wild''. More precisely, we run each API for 8 hours. The results show that developers confirmed the majority of the bugs (\textasciitilde$87\%$) we reported, suggesting that our mechanism for filtering false alarms was effective. Four of the bugs reported in \torch\ were rejected, either because the behavior was in accordance with the documentation or because the difference between CPU and GPU outcomes was acceptable according to developers, \ie{}, insufficient to characterize a bug and trigger a fix. Table~\ref{tab:bug_types} breaks down the confirmed bugs into six categories. We use a categorization similar to one used in recent prior work (e.g., \pathfinder~\cite{PathFinder_ICSE25}). These results suggest that \tname\ is capable of revealing a diverse range of bugs.

\begin{mdframed}[style=mpdframe,frametitle=\UseMacro{eq-in-the-wild}] \tname\
was highly effective in finding bugs, revealing \numBugs{}
previously-unknown bugs (\numBugsTorch\ in \torch\ and \numBugsTF\
in \tf) with different error manifestations.
\end{mdframed}

\subsection{Answering \EQ{3}:~\UseMacro{eq-ablation}}
\label{sec:ablation}
\begin{table*}[!tb]
\centering
\caption{Ablation study results for \tname. For each metric, the right-adjacent column ($\Delta$) shows the change relative to the full \tname configuration (FlashFuzz); arrows indicate increases ($\uparrow$) or decreases ($\downarrow$), and $\leftrightarrow$ denotes no change.}
\label{tab:ablation}
\setlength{\tabcolsep}{2pt}
\scriptsize
\begin{tabular}{l|rcrcrc|rcrcrc}
\toprule
\multirow{2}{*}{\textbf{Configuration}} & \multicolumn{6}{c|}{\textbf{PyTorch}} & \multicolumn{6}{c}{\textbf{TensorFlow}} \\
\cmidrule{2-7}\cmidrule{8-13}
& \textbf{Validity} & \textbf{$\Delta$} & \textbf{Coverage} & \textbf{$\Delta$} & \textbf{Crashes} & \textbf{$\Delta$} & \textbf{Validity} & \textbf{$\Delta$} & \textbf{Coverage} & \textbf{$\Delta$} & \textbf{Crashes} & \textbf{$\Delta$} \\
\midrule
FlashFuzz & 64.75 & \textemdash & 2501 & \textemdash & 2 & \textemdash & 92.89 & \textemdash & 1336 & \textemdash & 9 & \textemdash \\
w/o Documentation & 34.44 & \textcolor{red}{30.31\% $\downarrow$} & 1985 & \textcolor{red}{516 $\downarrow$} & 2 & {0 $\leftrightarrow$} & 61.45 & \textcolor{red}{31.44\% $\downarrow$} & 593 & \textcolor{red}{743 $\downarrow$} & 9 & {0 $\leftrightarrow$} \\
w/o Helper Functions & 77.04 & \textcolor{blue}{12.29\% $\uparrow$} & 1140 & \textcolor{red}{1361 $\downarrow$} & 0 & \textcolor{red}{2 $\downarrow$} & 76.00 & \textcolor{red}{16.89\% $\downarrow$} & 581 & \textcolor{red}{755 $\downarrow$} & 0 & \textcolor{red}{9 $\downarrow$} \\
w/o Doc+Helper Func & 23.70 & \textcolor{red}{41.05\% $\downarrow$} & 420 & \textcolor{red}{2081 $\downarrow$} & 0 & \textcolor{red}{2 $\downarrow$} & 53.55 & \textcolor{red}{39.34\% $\downarrow$} & 356 & \textcolor{red}{980 $\downarrow$} & 1 & \textcolor{red}{8 $\downarrow$} \\
\bottomrule
\end{tabular}
\end{table*}

To assess the contribution of each component of \tname, we performed an ablation
study in which we removed, one at a time, two elements from the LLM prompt: (i)
API documentation and (ii) helper functions. We did not ablate the template
component because it is required to compile harnesses, measure validity, and
handle errors. We evaluated the effect on three metrics: validity ratio, branch
coverage, and the number of crashes found. Results are summarized in
Table~\ref{tab:ablation}. For each framework (\torch and \tf), we constructed
the target set by combining APIs on which \tname previously revealed crashes
with a random sample from the supported API list, yielding 100 targets per
framework. We allocated a 10-minute budget per API. In Table~\ref{tab:ablation},
arrows indicate whether a metric increased (↑) or decreased (↓) relative to the
full \tname configuration.

The results show that documentation and helper functions affect the two
frameworks differently. For \tf, both components substantially aid validity and
coverage: removing documentation reduces validity from $92.89\%$ to $61.45\%$
and coverage from $1336$ to $593$ (crashes remain $9$); removing helper
functions also reduces validity (to $76.00\%$) and coverage (to $581$) and
eliminates crashes ($9\!\rightarrow\!0$). Removing both yields the lowest
validity ($53.55\%$) and coverage ($356$), with a single crash. For \torch,
documentation improves both validity and coverage: without documentation,
validity drops from $64.75\%$ to $34.44\%$ and coverage from $2501$ to $1985$
(crashes remain $2$). Removing helper functions increases validity (to
$77.04\%$) but substantially reduces coverage (to $1140$) and eliminates
crashes ($2\!\rightarrow\!0$). Overall, documentation consistently boosts
validity and coverage; helper functions trade higher coverage and crash
discovery for lower validity in \torch, but aid both validity and coverage in
\tf.

\begin{mdframed}[style=mpdframe,frametitle=\UseMacro{eq-ablation}]
Overall, results demonstrate that the use of documentation and helper functions
are essential for the performance we obtain using \libfuzzer\ with the test harnesses \tname\ generates.
\end{mdframed}

\section{Discussion}
\label{sec:discussion}

We discuss a small sample of bugs we find (Section~\ref{sec:bugs}), known limitations of this study (Section~\ref{sec:limitations}), threats to validity (Section~\ref{sec:threats}), and lessons learned (Section~\ref{sec:lessons}).

\subsection{Bugs detected by \tname}
\label{sec:bugs}

We describe below a few representative bugs that \tname\ detected in \torch\ and
\tf. The full list of bugs is available in our artifacts repository.

\MyPara{\textbf{\torch Bug
\href{https://github.com/pytorch/pytorch/issues/153337}{\#153337}}} PyTorch's
\CodeIn{torch.combinations} API computes $r$-length combinations of elements
from a 1D tensor. \tname{} detects a bug in this API, as shown by the code
snippet below that contains a fault-revealing input tensor (11 complex numbers)
and an API call that requests all length-10 combinations with replacement. This
input is valid according to the API documentation, but it causes the execution
to consume excessive memory and hang. 

\vspace{0.1ex}
\begin{lstlisting}[language=Python, basicstyle=\scriptsize,
  caption={Performance bug~\href{https://github.com/pytorch/pytorch/issues/153337}{\#153337} in PyTorch 2.7.0},
  captionpos=b,
  label={lst:pytorch-bug-1}]
tensor = torch.tensor(
    [
        -1.7267e-23 + 5.2018e-31j,
        # ... 9 intermediate complex values ...
        -1.2775e+17 - 5.5035e-17j,
    ], dtype=torch.cfloat,
)

torch.combinations(tensor, r=10, with_replacement=True) 
\end{lstlisting}

\emph{Root Cause:} The C++ backend implementation of this API
(\CodeIn{aten/src/ATen/native/Itertools.cpp}) builds the entire $r$-fold
Cartesian product using \CodeIn{at::meshgrid} and only then filters it. As a
result, this API allocates large intermediate tensors of size $n^r$ (data and
indices), ignoring the fact that the final output will only be of size
$\binom{n}{r}$ or $\binom{n+r-1}{r}$. Since $n^r$ grows extremely fast, the call
consumes a lot of memory and hangs. Note that this is a deep C++ backend design
bug; not a Python API misuse. Following our bug report, other users have also
reported similar issues in the same API, showing that this is a practical and
common problem.

\MyPara{\textbf{\tf Bug
\href{https://github.com/tensorflow/tensorflow/issues/94117}{\#94117}}} The
\tf API \CodeIn{tf.raw\_ops.ResourceSparseApplyProximalAdagrad} performs
sparse updates on entries in the input tensor variables according to the FOBOS
algorithm. The code snippet below attempts to update \CodeIn{var} and
\CodeIn{accum} using a learning rate \CodeIn{lr} and regularization parameters
\CodeIn{l1} and \CodeIn{l2}. The documentation does not specify the range or
type of input tensor \CodeIn{var} and \CodeIn{accum}, and \CodeIn{l1},
\CodeIn{l2}, and \CodeIn{lr} are scalars as the documentation expects.

\vspace{0.1ex}

\begin{lstlisting}[language=Python, basicstyle=\scriptsize,
  caption={Crash bug \href{https://github.com/tensorflow/tensorflow/issues/94117}{\#94117} present in \tf 2.19.0}, captionpos=b, label=lst:tf-bug-1]
large_val = 5.24393461e+36  # large magnitude to trigger crash
var   = tf.Variable([[large_val]*3, [large_val]*3], dtype=tf.float32, name="var")
accum = tf.Variable([[large_val]*3, [large_val]*3], dtype=tf.float32, name="accum")
lr = l1 = l2 = tf.constant(large_val, dtype=tf.float32)  # scalar hyperparameters
grad = tf.constant([7.90505e+31], dtype=tf.float32)  # single large gradient
indices = tf.constant([0], dtype=tf.int32)  # sparse update at index 0
tf.raw_ops.ResourceSparseApplyProximalAdagrad(var=var.handle, accum=accum.handle, lr=lr, l1=l1, l2=l2, grad=grad, indices=indices, use_locking=False)  # crashes with large values
\end{lstlisting}

However, when the input tensors \CodeIn{var} and \CodeIn{accum} contain large
values (in the order of $10^{30}$), they cause the API to crash with a fatal
error: \CodeIn{Check failed: d < dims() (1 vs. 1)}. Ideally, the API should
handle large input values gracefully without crashing. The developers have
confirmed that they can reproduce this bug using our input. However, it is still
pending a developer fix.


\MyPara{\textbf{\tf Bug
\href{https://github.com/tensorflow/tensorflow/issues/94379}{\#94379}}}
\tf API \CodeIn{tf.raw\_ops.BiasAdd} adds a 1D bias to the given
\CodeIn{value} tensor along the channel dimension specified by
\CodeIn{data\_format}. The snippet below triggers a GPU crash when
\CodeIn{value} is a 2D tensor with \CodeIn{data\_format = ``NCHW"}.

\vspace{0.1ex}

\begin{lstlisting}[language=Python, basicstyle=\scriptsize,
    caption=CPU/GPU inconsistency bug \href{https://github.com/tensorflow/tensorflow/issues/94379}{\#94379} (GPU crash in \CodeIn{BiasAdd} with 2D \CodeIn{value}),
    captionpos=b,
    label=lst:tf-bug-4]
# Crashed on GPU
with tf.device('/gpu:0'):
    value = tf.constant([0]*12, dtype=tf.int32, shape=[2, 6])
    bias = tf.constant([1, 2, 3, 4, 5, 6], dtype=tf.int32, shape=[6])
    result = tf.raw_ops.BiasAdd(value=value, bias=bias, data_format="NCHW")
\end{lstlisting}

The error is caused by passing a 2D tensor to the GPU implementation
with \CodeIn{data\_format = "NCHW"}; the GPU backend aborts with a
shape check failure (e.g., \CodeIn{Check failed: d < dims() (2 vs. 2)}),
while the CPU backend handles the same case silently, \ie{}, without raising any exception. The
documentation does not restrict the number of dimensions beyond
requiring rank $\ge 2$.

\subsection{Limitations}
\label{sec:limitations}

\MyPara{Accuracy of harnesses in generating valid inputs} The ability of our
method to produce valid inputs depends largely on (1) how accurately the LLM
interprets the API documentation to identify constraints, and (2) how accurately
it transforms raw bytes into valid inputs. We observe that LLMs capture these
requirements for a majority of harnesses, but we still see failures.
Quantitatively, for \torch, we enumerated 1,576 C++ candidate APIs: \tname built
1,151 (73\%) runnable harnesses that execute the API under test, while 340 (22\%)
failed to build due to compilation/linking errors, 72 (5\%) failed due to
missing target API, and 13 (0.8\%) failed at runtime. For \tf, among 1,452
candidate APIs, \tname built 662 (46\%) runnable harnesses that execute the API
under test, while 741 (51\%) failed due to compilation/linking errors, 43 (3\%) failed due to missing target API, and 6 (0.4\%)
failed at runtime.

We examined several failing harnesses. Out of ten randomly sampled harnesses
that failed to build or run, four failed to build because the target API was
missing in the backend, and six failed due to syntax errors or incorrect API
usage. In a separate random sample of ten missing target API cases, all were caused by
the LLM not following instructions and selecting a similar API instead. Four of
the runtime failures were silent, which are difficult for libFuzzer to detect.


\MyPara{Fuzzing on Backend vs FrontEnd} Since \tname\ focuses on fuzzing the C++
 backend, it may miss bugs that are present only in the Python frontend or the
 glue code between the frontend and backend. However, in \dl frameworks, the
 lion share of the code typically resides in the backend. Further, some
 frameworks like \tf have additional validity checks in the frontend that are
 not present in the backend. As a result, we observe that some potential bugs
 found by \tname{} turn out to be false positives when tested with the
 Python frontend. 

\MyPara{Performance Considerations}
We configured the fuzzer for maximum throughput. In particular, we capped
maximum input length at 128 bytes, limited mutation depth to 8, enabled two
threads to overlap I/O on a single core, and used a single in-process runner to
avoid TensorFlow/PyTorch initialization costs. These configurations are crucial
to maximize coverage and throughput. However, we did not explore all possible
optimizations. Further, we did not use any domain-specific mutators, which may
further improve coverage and bug-finding ability.

\subsection{Threats to Validity}
\label{sec:threats}

\MyPara{External Validity} A threat to the external validity of \tname\ concerns its
generalizability across different deep learning libraries. This threat arises
because our evaluation focuses primarily on \torch\ and \tf, which may limit the
applicability of our findings to other frameworks. However, \torch and \tf are
still the most widely used and most general deep learning libraries, and so our
findings may still be relevant to other emerging libraries that share similar
design principles, such as JAX~\cite{jax2018github}.

\MyPara{Internal Validity} An internal threat to validity is the reliability of
our LLM-based test harness generation. The quality of generated harnesses
depends on the LLM's ability to correctly interpret API documentation and
produce correct C++ test harness code. To mitigate this threat, we implemented a
multi-stage validation process for the generated harnesses. This process
includes compilation checks, manual inspection of a sample of harnesses, and
procedures to ensure that detected crashes originate from the target API rather
than the harness code itself.

\subsection{Lessons Learned and Future Directions}
\label{sec:lessons}

We list lessons that we learn from this study and directions for future work:

\begin{packed_itemize}
    \item \textbf{Coverage-guided fuzzing should be used in future evaluations
    of techniques for API-level fuzzing of \dl\ libraries.} Our
    results provided initial yet strong evidence that coverage-guided fuzzing is
    effective to find bugs in \dl\ APIs. Further, our results show that with
    systematic guidance, LLMs can generate correct harnesses even for complex
    domains like DL libraries. As a result, decades of research in
    coverage-guided fuzzers (like libFuzzer) can be brought to bear instead of
    developing fuzzers from scratch. Hence, we recommend the future work to use
    coverage-guided fuzzing as a (strong) baseline.
    
    \item \textbf{There is room to improve harness generation for \dl\ APIs.}
    While the LLM-based approach of \tname\ can generate harnesses for a large
    number of APIs (see Table~\ref{tab:num_target_apis}), there are still many APIs (425 for PyTorch, 790 for
    TensorFlow) we could not generate harnesses for. Furthermore, even when we
    generate harnesses, the coverage and validity ratio is low for some APIs.
    Recent LLM-based techniques like CKGFuzzer~\cite{xu2024ckgfuzzer} that uses
    knowledge graphs to generate fuzz harnesses can potentially be combined with
    \tname\ to further improve harness generation. We leave it for future work.

    \item \textbf{Triaging needs more attention.} We spent between 10-30
    minutes triaging each potential bug we found. After finding a likely bug
    using our methodology, we needed to check if the bug was unique by analyzing
    stack traces and then translate C++ inputs (byte arrays) that invoke the API
    backend operator into an input that invokes the corresponding API in Python
    (\textsection~\ref{tech:validation}). Translating that input manually is
    non-trivial and so we use an LLM for this task. But even doing so requires
    adaptations, for instance, when the tensors data is often too large to be
    fed to the LLM. In those cases, we manually curate a smaller representative tensor.
    Furthermore, sometimes the LLM is unable to produce correct Python code or
    the inputs require further minimization before we can report it to
    developers -- we manually solve such cases. To sum up, triaging failures is
    a very important part of the fuzzing process and automated debugging
    techniques would greatly benefit the community.

    
\end{packed_itemize}

\section{Related Work}
\label{sec:related}



\subsection{Test Harness Generation}
\label{sec:related-harness}
A few recent approaches such as CKGFuzzer~\cite{xu2024ckgfuzzer} incorporate code
knowledge graphs for harness generation, but obtaining those graphs for the high number of APIs on these libraries may not scale.
ElFuzz~\cite{chen2025elfuzz} uses LLMs to synthesize test
harnesses that directly embed syntactical and semantic constraints in the test
harness instead of synthesizing grammars or constraints separately. However,
their approach does not explicitly consider the structure of the API. 
Nexzzer~\cite{lin2025automatic} is a recent work that
extracts API relations and constraints from unit tests and static analysis to
generate API sequences for fuzzing. It is complementary to \tname\ as it focuses on properties of API sequences in contrast to properties of individual APIs. 
PromptFuzzer~\cite{lyu2024prompt} mutates LLM prompts based on coverage
feedback to generate diverse fuzz drivers, but it is not designed for
DL-specific data structures, like tensors. \tname\ is designed to handle the
unique challenges of testing \dl\ library APIs, such as parsing tensor representations and handling complex input constraints. We have yet to evaluate other methods of harness generation that may demonstrate even more favorable results for coverage-guided fuzzing.





\subsection{DL Library Testing}
\label{sec:related-dl-testing}



\subsubsection{Oracles}
\label{sec:related-diff}

Testing \dl\ libraries is challenging due to the lack of formal specifications.
Prior work relies on metamorphic and differential oracles to find bugs in \dl\ libraries.
Metamorphic oracles check whether certain properties hold under input transformations, as shown by Ding et al.~\cite{ding2017validating} and Xiao et al.~\cite{xiao2022metamorphic}.
Differential oracles compare outputs from different implementations or backends.
For example, CRADLE~\cite{Phan_ETAL_ICSE2019} compares results from CPU and GPU backends, while Li et al.~\cite{li2024dllens} use LLMs to translate APIs for cross-library differential testing.
Our study focuses on differential oracles that can reveal output inconsistencies and crashes; we leave the evaluation of API-specific metamorphic oracles for finding silent bugs to future work.

\subsubsection{Model-level fuzzing} 
\label{sec:model-level-fuzzing}

To address the limited test generation capabilities of earlier methods like
CRADLE, fuzzing has become a popular approach for thoroughly testing
\dl\ libraries. Model-level fuzzers use entire \dl\ models as test inputs. For
instance, AUDEE~\cite{AUDEE_ASE20} applies genetic algorithms to mutate
models—modifying inputs, weights, or both—and detects bugs via backend
inconsistencies or crashes. LEMON~\cite{LEMON_FSE20} modifies models at
the architectural level, while MUFFIN~\cite{MUFFIN_ICSE22} employs
neural architecture search (NAS) strategies on model computation graphs to find
bugs.
Model-level fuzzing also extends to \dl\ compilers. NNSmith~\cite{Liu_2023} uses
SMT solvers to generate valid models for testing compilers.
NeuRI~\cite{Liu_ETAL_ICSE23} infers operator rules from existing code
to create test models, proving effective for testing both compilers and,
consequently, \dl\ libraries. 
Despite their successes, model-level approaches can struggle with input
diversity due to strict inter-API constraints and may overlook less common APIs.
These limitations motivated the exploration of API-level fuzzing.

\subsubsection{API-level fuzzing}
\label{sec:related-api}
The most popular language to use DL libraries in is Python. The core
implementation of these libraries like \torch\ and \tf\ are, however,
implemented in C++. Thus, two types of API-level fuzzers are found in the
literature: fuzzers on the Python frontend and fuzzers on the C++ backend.

\MyItPara{Fuzzing on the frontend}
\label{sec:related-frontend}
API-level fuzzing on the Python frontend is challenging because argument types
are not explicit. FreeFuzz~\cite{Wei2022} solves this problem by mining
open-source code snippets and executing them on instrumented DL libraries to
infer argument types and construct a value space for each API.
This approach achieves better code coverage than model-level fuzzers like LEMON.
Limitations in inferring constraints from frontend Python code
can be addressed by using API documentation, which often contains the majority
of constraints for specific DL library APIs. DocTer~\cite{Xie2022DocTer} uses
API-documentation to automatically infer constraints such as type information,
tensor shapes, and valid value ranges. While less comprehensive than
backend-inferred constraints (Section~\ref{sec:related-backend}), this work is
the first to use documentation-inferred constraints for DL library testing.
DeepREL~\cite{Deng_ETAL_FSE22} leverages similar signatures shared by different
DL Library APIs with relational API inference, allowing inputs to be reused for
multiple APIs.
Using this strategy, DeepREL covers more APIs than FreeFuzz and reveals numerous
bugs in DL libraries.

\ttfuzz~\cite{Deng_ETAL_ISSTA23} utilizes the generative LLM
Codex~\cite{chen2021evaluating} to create candidate seed programs based on
sequences of API invocations from DL libraries. These seeds are then refined
through evolutionary fuzzing and new programs are generated using the infilling
LLM InCoder~\cite{fried2022incoder}.
Similarly, FuzzGPT~\cite{deng2023large} employs Codex~\cite{chen2021evaluating}
and CodeGen~\cite{nijkamp2022codegen} to generate new code from bug reports and
partial or complete code snippets from DL libraries. This approach emulates
model-like behavior by generating code snippets with API sequences.

Despite the effectiveness of these tools in finding bugs, computing code coverage 
from the library frontend can be expensive. Hence, \tname\ uses efficient in-process coverage collection for
backend API-level fuzzing, improving both speed and effectiveness.

\MyItPara{Fuzzing on the backend}
\label{sec:related-backend}
\tname\ targets the backend of \dl\ libraries. One advantage of fuzzing the backend is that C++ is
strongly typed, unlike Python. IvySyn~\cite{Christou_ETAL_USENIX23}
takes advantage of this fact by using type-aware mutation in the backend code.
IvySyn also uses the existing frontend-to-backend code mapping to
convert crash-inducing low-level (C++) inputs into high-level (Python) inputs
that can trigger the bug via frontend API calls. However, this tool operates as
a black box, being unaware of the code executed within the API.

Since the core implementation of DL library APIs is in the backend, gray-box and
even white-box techniques become feasible. However, applying white-box
techniques such as symbolic execution~\cite{cadar2013symbolic,baldoni2018survey}
directly to the backend is computationally expensive due to the inherent
complexity of DL libraries. \pathfinder~\cite{PathFinder_ICSE25} employs a middle
ground solution by using inductive program synthesis to infer approximate paths.
\pathfinder\ uses lightweight instrumentation to collect branch coverage from test
inputs and applies program synthesis (using Duet~\cite{lee2021combining}) to infer
approximate path constraints for fuzzing. 
%
ACETest~\cite{Shi_2023} extracts constraints by tracing error-handling code in
the backend to user-controllable parameters, then uses an SMT
solver~\cite{de2008z3} for input generation, achieving promising results.
However, unlike \tname\ they do not use coverage-guidance, which limits their
effectiveness and efficiency. We observe that \tname{}'s test harnesses can
effectively capture API input constraints from API documentation in natural
language, allowing it to generate many valid inputs.



\section{Conclusions}
\label{sec:conclusions}

Finding bugs in Deep Learning (\dl) libraries is an important and challenging
problem. API-level fuzzing is a prominent technique for finding bugs in these libraries. Despite the impressive advances in techniques, the impact of coverage guidance on input data generation has not been studied in this domain. This paper reports the first in-depth study on using coverage-guided fuzzing (\cgf) --which has been successful in many other domains-- for testing \dl\ library APIs.

An important challenge in
using \cgf\ for API-level testing relates to scaling the creation of
test harnesses to handle APIs with various signatures and input
constraints. To enable this study, we propose an LLM-based technique, dubbed \tname, to obtain test harnesses for APIs using API documentation, a harness template, and helper functions. 
\tname\ wraps the coverage-guided fuzzer \libfuzzer~\cite{libfuzzer} with the capability of generating harnesses for \dl\ APIs.

Results show that \tname\ outperforms three \sota\ recently-proposed techniques (namely, \acetest,
\ttfuzz, and \pathfinder) in terms of coverage, validity rate, and ability to reveal crashes. A large-scale
study further demonstrates the practical utility of \tname, revealing \textbf{\numBugs}
previously unknown bugs in two recent versions of \torch\ and \tf. We learned several lessons from this study. In particular, we recommend --based on the experimental results-- that future research in API-level fuzzing use \cgf\ as a comparison baseline.

\section*{Data Availability}

The artifacts -- including datasets and scripts -- are publicly available~\cite{ourrepo}.

\bibliographystyle{ACM-Reference-Format}
\begingroup\catcode`\_=12 \relax
\bibliography{ref}
\endgroup

\end{document}